\documentclass[12pt,preprint]{aastex}

\usepackage{amsmath}
\usepackage{graphics,graphicx}
\usepackage{subfig}
\usepackage{multirow}
\usepackage{array}
\usepackage{rotating}
\usepackage[T1]{fontenc}
\usepackage{textcomp}
\usepackage[hyphens]{url}
\usepackage{hyperref}

\begin{document}
%

\title{Analytic Scattering and Refraction Models for Exoplanet Transit Spectra}

\author{Tyler D. Robinson\altaffilmark{1,2}}
\affil{Department of Physics and Astronomy, Northern Arizona University, Flagstaff, AZ 86011, USA}
\email{tyler.robinson@nau.edu}

\author{Jonathan J. Fortney\altaffilmark{1}}
\affil{Department of Astronomy and Astrophysics, University of California, Santa Cruz, CA 95064, USA}

\and

\author{William B. Hubbard}
\affil{Lunar and Planetary Laboratory, University of Arizona, Tucson, AZ 85721-0092, USA}

\altaffiltext{1}{University of California, Santa Cruz, Other Worlds Laboratory}
\altaffiltext{2}{NASA Astrobiology Institute's Virtual Planetary Laboratory}

%
\begin{abstract}

Observations of exoplanet transit spectra are essential to understanding the physics and 
chemistry of distant worlds.  The effects of opacity sources and many physical processes 
combine to set the shape of a transit spectrum.  Two such key processes---refraction and 
cloud and/or haze forward scattering---have seen substantial recent study.  However, models 
of these processes  are typically complex, which prevents their incorporation into observational 
analyses and standard transit spectrum tools.  In this work, we develop analytic expressions 
that allow for the efficient parameterization of forward scattering and refraction effects in 
transit spectra.  We derive an effective slant optical depth that includes a correction 
for forward scattered light, and present an analytic form of this correction.  We validate 
our correction against a full-physics transit spectrum model that includes scattering, and 
we explore the extent to which the omission of forward scattering effects may bias models.  
{ Also, we verify a common analytic expression for the location of a refractive boundary, 
which we express in terms of the maximum pressure probed in a transit spectrum.  This 
expression is designed to be easily incorporated into existing tools, and we discuss how 
the detection of a refractive boundary could help indicate the background atmospheric 
composition by constraining the bulk refractivity of the atmosphere.}  Finally, we show that 
opacity from Rayleigh scattering and collision induced absorption will outweigh the effects of 
refraction for Jupiter-like atmospheres whose equilibrium temperatures are above 400--500~K.

\end{abstract}
%


%
\section{Introduction}

Transit spectroscopy \citep{seager&sasselov2000,brown2001,hubbardetal2001} 
is an important technique used in the study of exoplanet atmospheric composition.  
Observations of transit spectra are now a common approach to studying a diversity of planet 
types with NASA's {\it Hubble Space Telescope} \citep{fraineetal2014,knutsonetal2014b,
kreidbergetal2014b,singetal2016}.  Exoplanet transit spectroscopy will also be a 
key science component of NASA's {\it James Webb Space Telescope} ({\it JWST}) 
\citep{gardneretal2006}, which is expected to provide high-quality transit spectra of 
many tens of targets over the duration the mission \citep{beichmanetal2014}.

Recently, certain physical processes that were omitted from early models of exoplanet 
transit spectra have been shown to be of significant importance in certain circumstances. 
In particular, atmospheric refraction can lead to ``floors'' in transit spectra of worlds with 
relatively transparent atmospheres, or worlds that are located far from their host star so 
that the angular size of the host star as seen from the planet is small \citep{sidis&sari2010,
munozetal2012,betremieux&kaltenegger2014,misraetal2014,betremieux&kaltenegger2015,
dalbaetal2015,betremieux2016}.  Additionally, forward scattering from atmospheric 
clouds and hazes has been shown to be an important consideration for exoplanets 
whose host star is of relatively large angular size as seen from the planet 
\citep{dekok&stam2012,robinson2017}.

Incorporating refraction and/or scattering into a transit spectrum model typically 
comes at great computational cost.  For the former, ray tracing simulations  
need to be implemented, and the latter warrants three-dimensional Monte Carlo 
techniques.  Thus, typical transit tools, which place a strong emphasis on 
computational efficiency, neglect the physics of refraction or scattering 
\citep[e.g.,][]{lineetal2011,benneke&seager2012,howe&burrows2012,
barstowetal2012,leeetal2014,waldmannetal2015,morleyetal2016}.

Given the desire to balance realistic physics with computational efficiency in 
transit spectrum models, there is a clear need for the development of fast approaches 
to incorporating both refraction and scattering effects into standard modeling tools.  
Also, with the improvement in observational precision expected from {\it JWST}, 
now is an opportune time to examine the fundamentals of the physics that shape 
transit spectra.  { Here, we use simple physical arguments to derive an analytic 
correction that accounts for forward scattering effects in transit, and we use 
previously-published expressions to arrive at a simple equation for the pressure level 
of the refractive boundary for a transiting planet.}  These corrections 
are designed to be easily incorporated into any standard transit spectrum model, and can 
also be used to understand the basic physics of refraction and scattering in transit spectra.
To ensure validity, we compare results from key expressions against those from more 
sophisticated techniques.  Finally, we use { the refractive boundary treatment} to explain 
the circumstances under which refraction is likely to be an important process in shaping 
the transit spectra of worlds whose atmospheres are primarily H$_2$ and He.

\section{Theory}
\label{sec:theory}

In transit spectroscopy, the fundamental consideration is whether or not rays that 
traverse an exoplanet atmosphere during transit connect the observer to the stellar 
disk.  A patch on the disk of a transiting exoplanet will appear opaque if rays 
emerging from this location are strongly attenuated by gas or aerosol absorption 
opacity, or if these rays experience enough refraction to bend them off the host 
stellar disk.  Scattering opacity can also cause the patch to become opaque, since 
aerosols that are not strongly forward scattering are likely to direct rays away from 
the host stellar disk.  Considering all of these processes, it is apparent that the 
key quantities that determine the transmitted light emerging from a patch on the disk 
of a transiting exoplanet are: the slant absorption optical depth, $\tau_{\rm{a}}$, the 
slant scattering optical depth, $\tau_{\rm{s}}$, the normalized scattering phase function, 
$P(\theta)$ (where $\theta$ is the scattering angle), the refraction angle, $\omega$, 
the projected separation between the patch and the opposite limb of the stellar host, 
and the angular size of the host star as seen from the planet.  A number of these 
quantities depend on wavelength, and we leave this dependence implicit for cleaner 
presentation.  

The following subsections present simple models that enable scattering 
and refraction effects to be easily incorporated into standard transit spectrum tools.  We 
derive these using the typical assumption that the entire planetary disk overlays a 
star of uniform surface brightness, as limb darkening is accounted for in standard transit 
observation data reduction procedures \citep[e.g.,][]{mandel&agol2002,kreidberg2015}.  
We also assume that the planet is centered on the stellar disk, as this is the most 
straightforward regime for computing transit spectra, although we note pieces of our 
theory that can be easily generalized for cases that break our adopted symmetry.  
Finally, our theory will not account for brightening effects related to refraction 
\citep{misra&meadows2014} or forward scattering \citep{robinson2017} while the 
planet is slightly off or partially overlapping the stellar disk.

\subsection{Scattering Effects}
\label{subsec:scat}

Standard transit spectrum models treat all optical depth (i.e., the combined absorption 
and scattering optical depths) as absorption optical depth.  However, the forward scattered 
portion of a ray will still be directed towards the observer.  Thus, an improved 
treatment of attenuation in transit spectra would reduce the scattering optical depth by 
an amount that depends on the extent of the forward scattering peak in the scattering 
phase function, and then equate the remaining scattering and absorption optical 
depths with the effective absorption optical depth.

{ To begin, we adopt the $\delta$-isotropic approximation to radiative transfer 
\citep[][their Section~6.8.1]{thomas&stamnes1999}, where the scattering phase function is 
represented by an isotropic component and a forward component of strength $f$ 
(with $0 \leq f \leq 1$).  Here, the radiant intensity, $I$, along a path (which we imagine 
as the slant path) would obey,
\begin{equation}
  \frac{dI}{d\tau_{\rm{e}}} = - I + f \tilde{\omega}_{0} I + \frac{\tilde{\omega}_{0}\left(1-f\right)}{4\pi} \int_{4\pi} I\left(\Omega^\prime\right) d\Omega^\prime \ ,
\label{eqn:deltarte}
\end{equation}
where $\tau_{\rm{e}}$ is the extinction optical depth (i.e., the sum of the absorption and 
scattering optical depths), $\tilde{\omega}_{0}$ is the single scattering albedo, and $\Omega$ 
represents a solid angle.  The final term in Equation~\ref{eqn:deltarte} comes from the 
isotropic component of the phase function, which has been shown to be negligible for transit 
spectra \citep{brown2001,hubbardetal2001,robinson2017}, so we choose to omit this term.  
With these assumptions, the radiative transfer equation can be written simply as 
\begin{equation}
  \frac{dI}{d\tau_{\rm{eff}}} = -I \ ,
\end{equation}
where we have defined an effective optical depth that includes a correction for forward 
scattering as,
\begin{equation}
  d\tau_{\rm{eff}} = (1 - f\tilde{\omega}_{0}) d\tau_{\rm{e}} \ .
\label{eqn:taueff}
\end{equation}}

{ It is widely known in the radiative transfer literature \citep[e.g.,][their Appendix A.2]{petty2006} 
that there is no unique way to set the value of the forward scattering correction factor.  Given the 
geometry of the problem at hand, in which we imagine back-tracing rays from an observer into the 
planetary atmosphere, where these rays then scatter into a cone of radiation that either fully or 
partially overlaps the host stellar disk,} we suggest that the corrective factor should be based on the 
portion of the scattering phase function that is within the angular size of the stellar disk.  This 
would then give,
\begin{equation}
  f = 2\pi \int_{0}^{\frac{R_{\rm{s}}}{a}} P(\theta) \sin\left(\theta\right) d\theta = 2\pi \int_{1 - \frac{R_{\rm{s}}}{a}}^{1} P(\mu) d\mu \ ,
\label{eqn:fcorr}
\end{equation}
where $R_{\rm{s}}$ is the stellar radius, $a$ is the planet-star physical separation, 
$\mu = \cos(\theta)$, and we have assumed that $\sin(R_{\rm{s}}/a) \approx R_{\rm{s}}/a$.

{ In practice, Equation~\ref{eqn:taueff} would simply be used in place of the extinction 
optical depth in any standard transit spectrum model.  Such models typically operate by integrating 
the optical depth along a slant path, which yields a transmissivity through the atmosphere for a 
given impact parameter.  In such a setup, a layer along the slant path that has any scattering 
optical depth (i.e., $\tilde{\omega}_{0} > 0$) would have the slant extinction optical depth 
across the layer reduced by $1 - f\tilde{\omega}_{0}$.  Thus, our proposed correction is 
computationally inexpensive to implement inside of existing tools.}

\begin{figure}
  \centering
  \includegraphics[trim = 0mm 0mm 0mm 0mm, clip, width=6.2in]{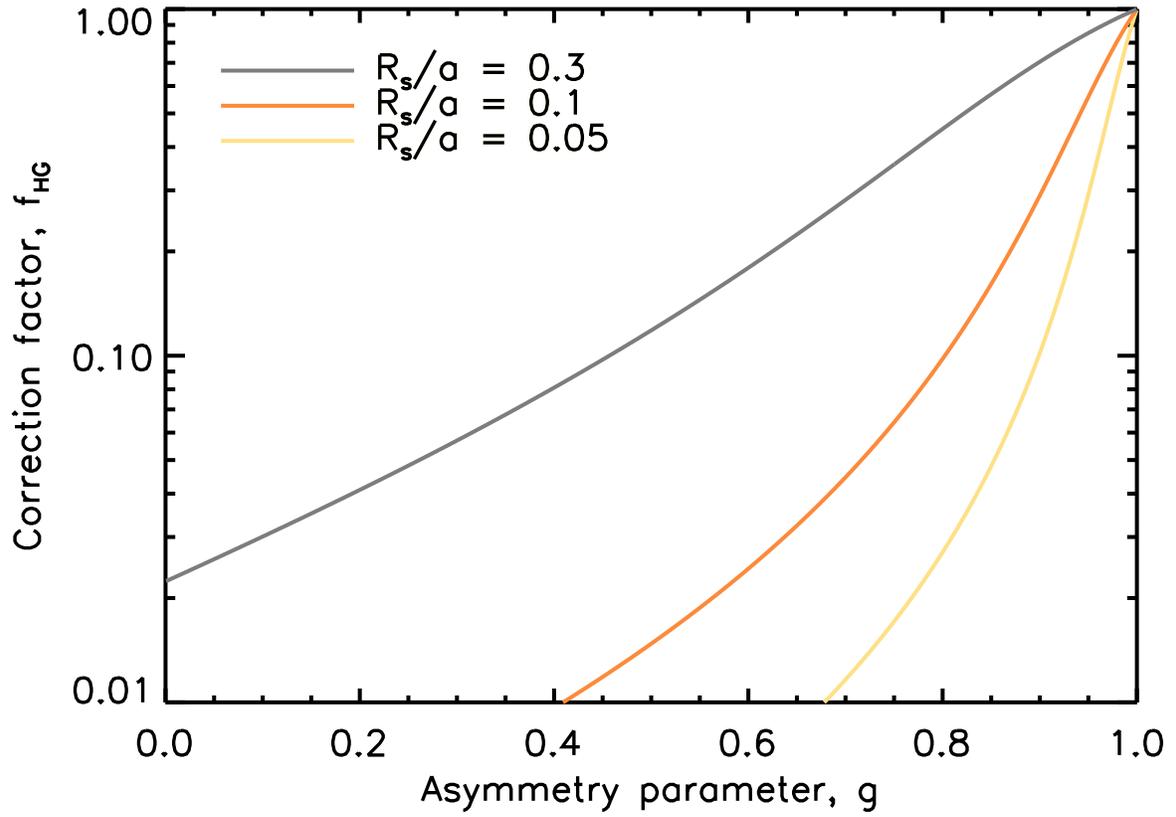}
  \caption{The forward scattering correction factor, assuming a Henyey-Greenstein phase 
                function, as a function of asymmetry parameter, $g$, for several different host star 
                angular sizes.}
  \label{fig:fHG}
\end{figure}

In the most general case, the integral in Equation~\ref{eqn:taueff} would 
need to be computed for each layer given a realistic scattering phase function.  This 
computational burden is reduced if the scattering phase function is expanded in terms 
of Legendre polynomials.  However, given that our correction factor is approximate, such 
sophistications are unwarranted.  We propose an analytic correction factor that is based on 
the widely used Henyey-Greenstein phase function \citep{henyey&greenstein1941}, with,
\begin{equation}
  P(\theta) = \frac{1}{4\pi} \frac{1 - g^2}{\left[ 1 + g^2 - 2g\cos\left(\theta\right) \right]^{3/2}} \ ,
\end{equation}
where $g$ is the asymmetry parameter.  The correction factor for this phase function is 
then,
\begin{equation}
  f_{\rm{HG}}\!\left(g,R_{\rm{s}}/a\right) = \frac{1 - g^2}{2g}\left[ \frac{1}{1 - g} - \frac{1}{\sqrt{1 + g^2 - 2g\cos\left(\frac{R_{\rm{s}}}{a}\right)}} \right] \ .
\label{eqn:fHG}
\end{equation}
Additional complexity could be added with little computational cost by using a two-term 
Henyey-Greenstein phase function \citep{irvine1965}, whose correction factor would be,
\begin{equation}
  f_{\rm{TTHG}}\!\left(g_{1},g_{2},w,R_{\rm{s}}/a\right) =  w \cdot f_{\rm{HG}}\!\left(g_{1},R_{\rm{s}}/a\right) + (1-w) \cdot f_{\rm{HG}}\!\left(g_{2},R_{\rm{s}}/a\right)\ ,
\end{equation}
where $g_{2} < 0$ causes an enhancement in backward scattering, and $w$ (whose 
value is between zero and unity) is a weighting factor for the forward versus backward 
scattering terms.  { Figure~\ref{fig:fHG} demonstrates the scale of the correction factor from  
Equation~\ref{eqn:fHG} as a function of the asymmetry parameter and for several different 
host star angular sizes.}  When adopting the correction appropriate for the Henyey-Greenstein 
phase function, the correction factor only varies as rapidly as does the asymmetry parameter.  

\begin{figure}
  \centering
  \includegraphics[trim = 0mm 0mm 0mm 0mm, clip, width=6.2in]{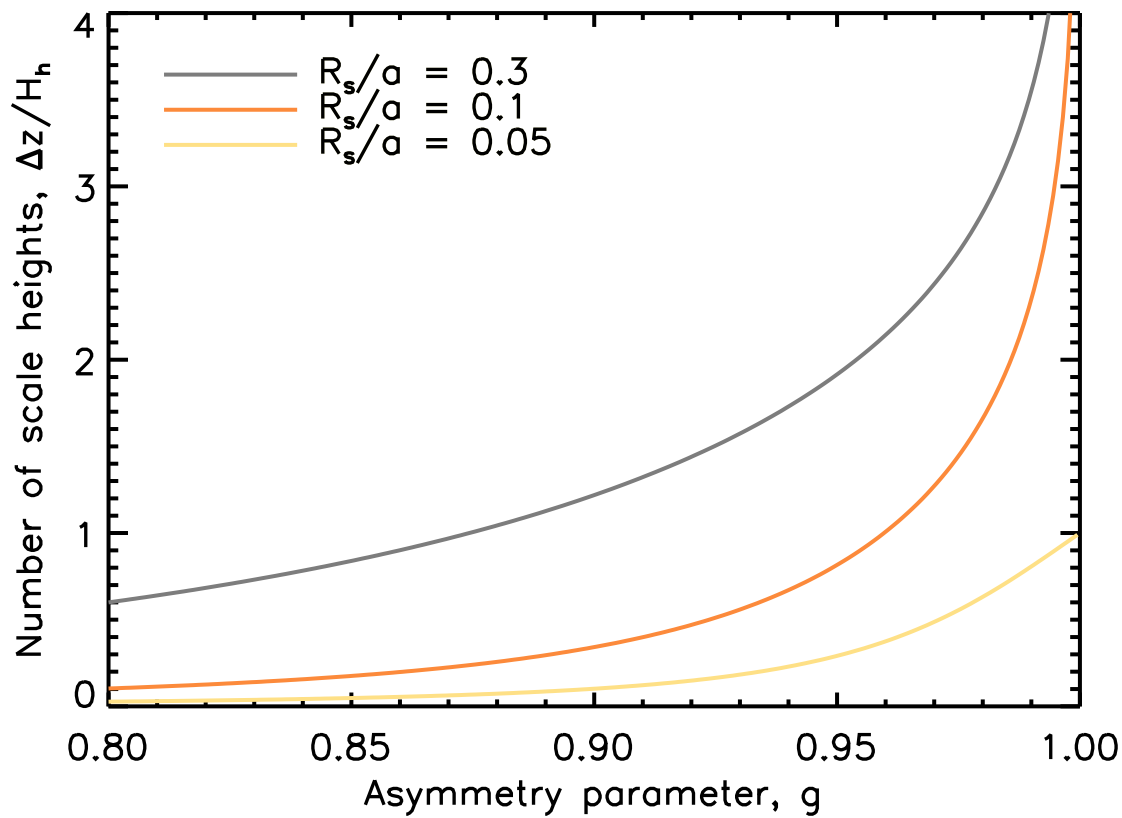}
  \caption{Difference in the altitude probed when our forward scattering correction is included versus 
                excluded, plotted relative to the haze scale height and taken from Equation~\ref{eqn:dzH}.  
                We assume a purely scattering haze, and adopt a Henyey-Greenstein phase 
                function as in Figure~\ref{fig:fHG}.  Curves are for different host star angular sizes.}
  \label{fig:dzH}
\end{figure}

{ Briefly, we note that, for an atmosphere where the dominant opacity source is an aerosol that is 
distributed vertically with some scale height, $H_{\rm{h}}$ (where the sub-script `h' is for `haze'), 
we can use Equation~8 from \citet{robinsonetal2014a} to derive a radius (or altitude) difference 
probed when our correction is included versus omitted.  The transit radius, $r$, would be set where  
the slant optical depth is of order unity, and would satisfy,
\begin{equation}
  1 \approx \tau_{0} \sqrt{ \frac{2\pi r}{H_{\rm{h}}} } e^{-\left(r-R_{\rm{p}}\right)/H_{\rm{h}}} \ ,
\end{equation}
where $\tau_{0}$ is a reference vertical extinction optical depth, $R_{\rm{p}}$ is the planetary 
radius, and we have omitted the wavelength dependence from the original expression.  Replacing 
$\tau_{0}$ with an optical depth that includes our forward scattering correction (i.e., 
$\left[1 - f\tilde{\omega}_{0}\right]\tau_{0}$) would produce a similar expression for the transit radius 
probed when forward scattering is considered, which we will call $r^{\prime}$.  Taking the ratio of 
these two expressions and solving for the radius difference relative to the haze scale height 
gives,
\begin{equation}
  \frac{r - r^{\prime}}{H_{\rm{h}}} = \frac{\Delta r}{H_{\rm{h}}} = - \ln\left(1 - f\tilde{\omega}_{0} \right) \ ,
\label{eqn:dzH}
\end{equation}
where the have assumed that $\Delta r \ll r$.  For these same angular sizes of the host star as 
in Figure~\ref{fig:fHG}, and for the adopted Henyey-Greenstein phase function, Figure~\ref{fig:dzH} 
demonstrates  Equation~\ref{eqn:dzH}, assuming pure scattering and with $\tilde{\omega}_{0}=1$.
The range of scale height biases shown in Figure~\ref{fig:dzH} (i.e., $\lesssim 4$) agrees with 
quoted results from the Monte Carlo scattering model of \citet{dekok&stam2012}.}

\subsection{Refraction Effects}

Following \citet{hubbardetal1988}, we derive the total refractive bending angle by 
determining the phase change, $\Phi$, of a wave passing on a straight-line path through 
the exoplanet atmosphere.  Since the refractivity, $\nu$, of the atmosphere is 
proportional to the number density, and assuming these to both be distributed 
exponentially in an isothermal atmosphere with scale height $H$, we have,
\begin{equation}
  \Phi = k \nu_{0} H \sqrt{\frac{2 \pi r}{H}} e^{-\left( b - r_{0} \right)/H} \ ,
\end{equation}
where $k$ is the wavenumber, $\nu_{0}$ is the refractivity evaluated at a radius 
$r_{0}$ in the planetary atmosphere, and $r$ is the radius of closest approach for the 
ray (which is equivalent to the impact parameter for our assumed straight-line path).  
The total refractive bending angle (taken to be non-negative) is then,
\begin{equation}
  \omega = -\frac{1}{k} \frac{\partial \Phi}{\partial r} = -\nu_{0} \left( \frac{1}{2} \sqrt{\frac{2\pi H}{r}} - \sqrt{\frac{2 \pi r}{H}} \right) e^{-\left( r - r_{0} \right)/H} \ .
\end{equation}
Since $r \gg H$, we have,
\begin{equation}
  \omega \approx \nu_{0} \sqrt{\frac{2 \pi r}{H}} e^{-\left( r - r_{0} \right)/H} \ ,
\label{eqn:refract}
\end{equation}
{ which is an expression first derived by \citet[][their Equation~4]{fabry1929}, and 
was also derived by \citet[][their Equation~6]{baum&code1953}.}

{\citet{sidis&sari2010} noted that the base of the transit spectrum (i.e., the smallest 
planetary radius probed, $r_{\rm{min}}$) would be determined by 
Equation~\ref{eqn:refract} when the refraction effect is large enough to cause a ray coming 
from one side of the planet to bend off the far limb of the star, which would give,
\begin{equation}
  \frac{R_{\rm{s}}+R_{\rm{p}}}{a} \approx \nu_{0} \sqrt{\frac{2 \pi r_{\rm{min}}}{H}} e^{-\left( r_{\rm{min}} - r_{0} \right)/H} \ .
\label{eqn:angfull}
\end{equation}
Evaluating this expression yields the minimum impact parameter (or radius, equivalently) 
probed \citep[][their Equation~17]{sidis&sari2010}.  \citet{betremieux&kaltenegger2014}, who 
call the bending angle highlighted by \citet{sidis&sari2010} the ``critical deflection'' angle, 
express Equation~\ref{eqn:angfull} in terms of the maximum atmospheric number density 
probed (in amagats; their Equation~11).

Instead of expressing the maximum depth probed in terms of number density or 
radius, we propose that the more intuitive measure is atmospheric pressure, $p$.  Taking 
$p=p_0 \exp\left[-(r - r_0)/H \right]$, where $p_{0}$ is the reference pressure at $r_{0}$, 
we can use the result of \citet{sidis&sari2010} to determine the maximum pressure 
probed by the ray, $p_{\rm{max}}$, as, 
\begin{equation}
  p_{\rm{max}} = p_0 \frac{1}{\nu_0} \frac{R_{\rm{s}} + R_{\rm{p}}}{a} \sqrt{ \frac{H}{2\pi r_0} } \approx p_0 \frac{1}{\nu_0} \frac{R_{\rm{s}}}{a} \sqrt{ \frac{H}{2\pi r_0} } \ ,
\label{eqn:pmax}
\end{equation}
where the approximation assumes $R_{\rm{p}} \ll R_{\rm{s}}$.}
For easy use, we can adopt Jupiter-like values and write, 
\begin{equation}
\begin{split}
  p_{\rm{max}} & = 23~{\rm{mbar}} \cdot \left( \frac{1.23\times10^{-4}~{\rm{(STP)}}}{\nu_0} \right) 
                              \left( \frac{T}{130~{\rm{K}}} \right)^{\!\!3/2} \left( \frac{R_{\rm{s}}}{R_{\odot}} \right) \\ 
    \times          &    \left( \frac{5.2~{\rm{au}}}{a} \right) \left( \frac{R_{\rm{J}}}{R_{\rm{p}}} \right)^{\!\!1/2}
                               \left( \frac{2.2~{\rm{amu}}}{\mu} \right)^{\!\!1/2}  \left( \frac{24.8~{\rm{m~s^{-2}}}}{g} \right)^{\!\!1/2} \ ,
\end{split}
\end{equation}
where $R_{\rm{p}}$ is taken as the 1~bar planetary radius, $\mu$ is the mean molar 
weight for the atmosphere, $g$ is the acceleration due to gravity, and the refractivity is 
given at Standard Temperature and Pressure (STP; 273~K and 1.013~bar).  { Note that 
including the planetary radius on the right hand side of Equation~\ref{eqn:pmax} serves to 
increase the maximum pressure probed by a factor of $R_{\rm{p}}/R_{\rm{s}}$, and this omission 
is unlikely to be a limiting factor (as compared to, e.g., the isothermal assumption).  Omitting the 
planetary radius contribution is justified as exoplanet transit depths tend to be smaller than 
2--3\% (implying $R_{\rm{p}}/R_{\rm{s}}$ less than 0.2)\footnotemark[1], and, furthermore, 
radial velocity results indicate that massive planets are relatively rare around low mass stars 
\citep{johnsonetal2010}.}  Table~1 contains expressions for computing the 
wavelength-dependent refractivity at STP for a number of key gases.  

\footnotetext[1]{{See: \url{exoplanetarchive.ipac.caltech.edu}}}

In Equation~\ref{eqn:pmax}, we note that the only term that depends on wavelength is 
the refractivity (which tends to vary weakly with wavelength across the near-infrared and 
thermal infrared).  Also, for cases that aim to compute transit spectra that break our assumed 
symmetry where the planet centered on the stellar disk, Equation~\ref{eqn:angfull} would need 
to be evaluated for a grid of locations around the planetary disk.  Each angular point would 
have a different refractive boundary pressure, and an area weighted average could be 
performed for the grid (in angle) of transit spectra for the planet.

\section{Results}

The scattering correction model and the refraction expression given above are easily implemented 
in transit spectrum tools.  Here we verify these simple expressions  
via comparisons to previously published analyses and to more sophisticated models.  We 
conclude this section by using the refraction model to understand the conditions under 
which refractive bending of light is likely to be important for atmospheres of Jupiter-like planets.

\subsection{Model Verifications}

\begin{figure}
  \centering
  \begin{tabular}{c}
    \includegraphics[trim = 0mm 0mm 0mm 0mm, clip, width=3.1in]{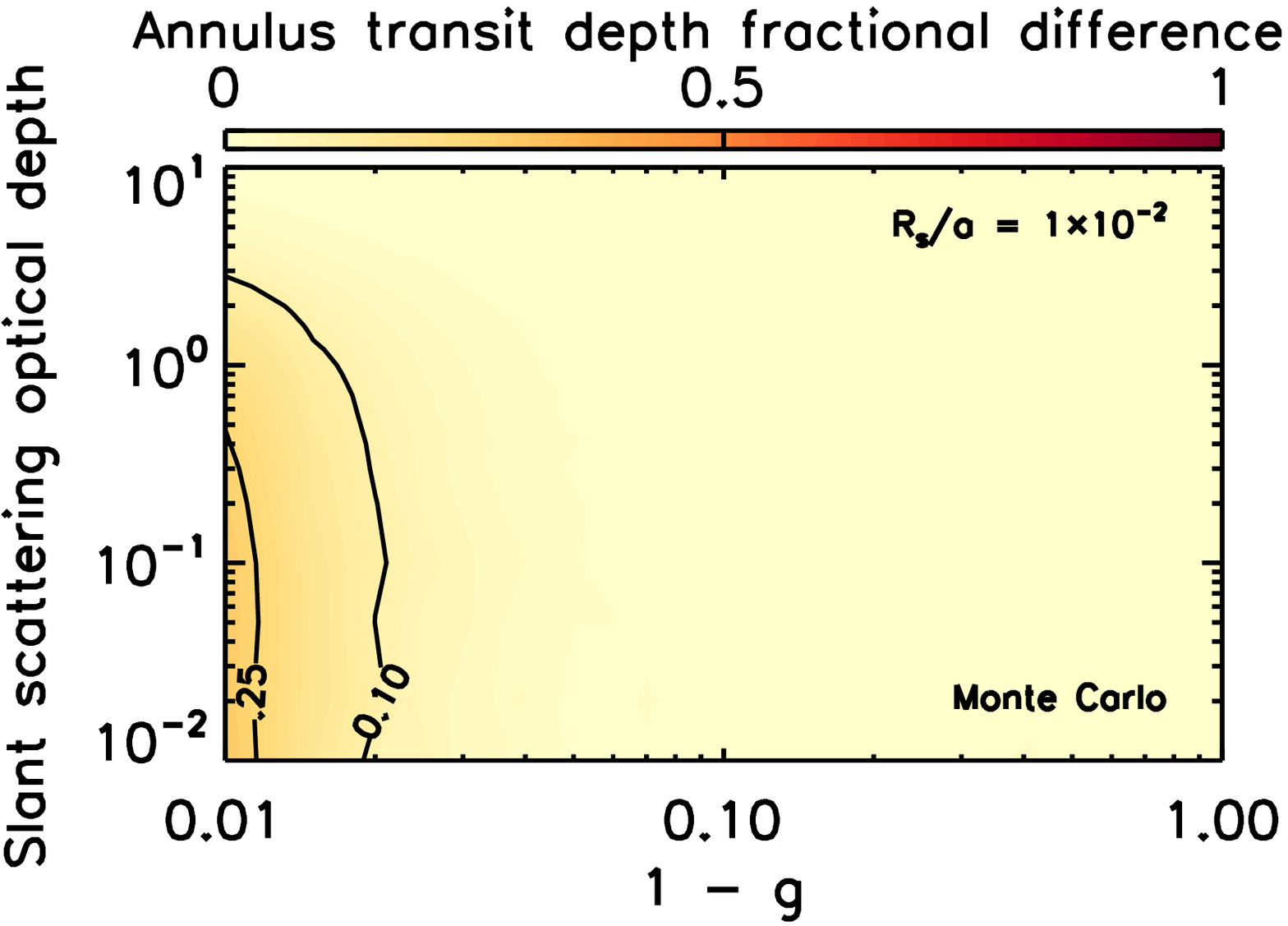}
    \includegraphics[trim = 0mm 0mm 0mm 0mm, clip, width=3.1in]{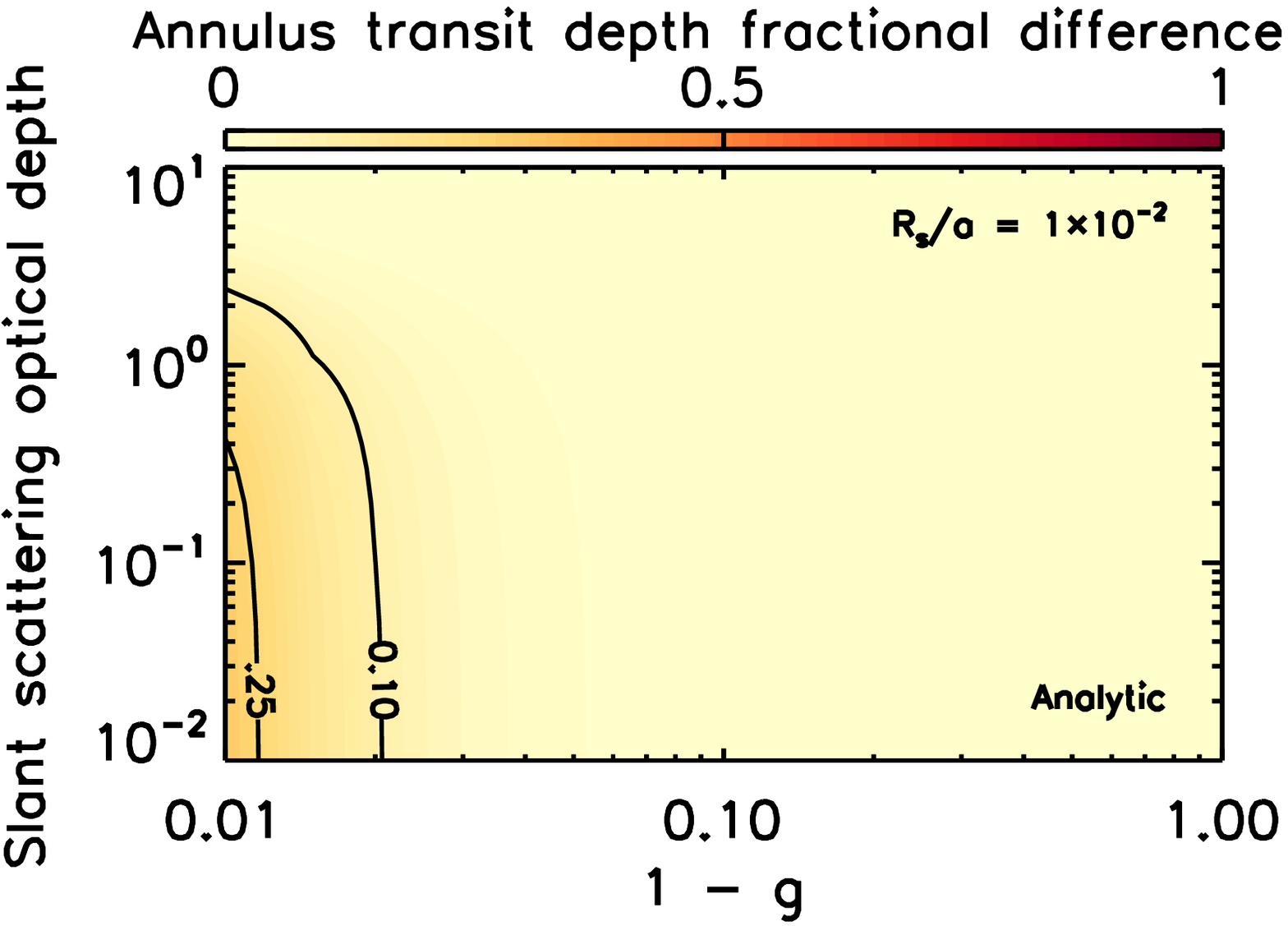} \\
    \includegraphics[trim = 0mm 0mm 0mm 0mm, clip, width=3.1in]{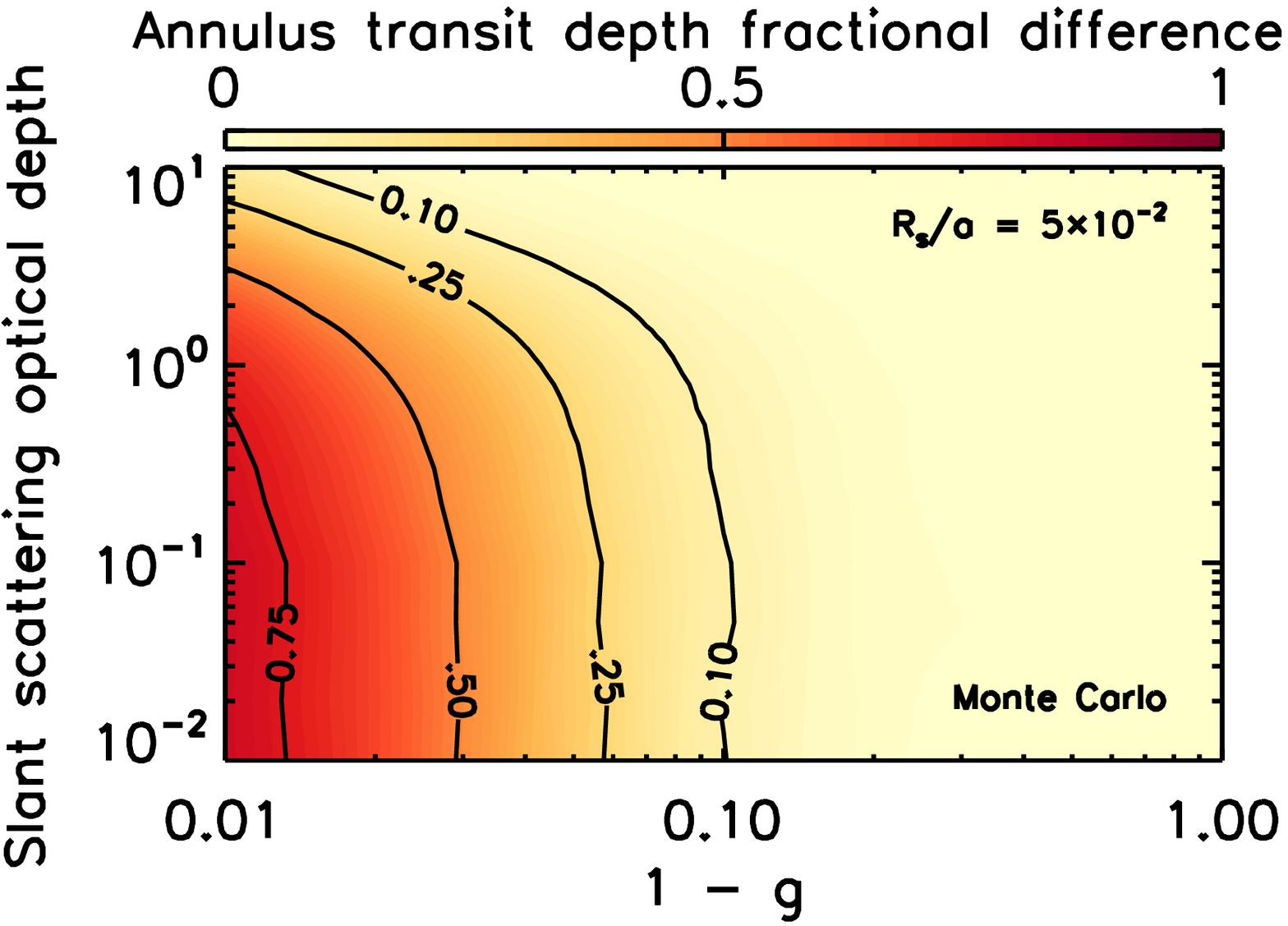}
    \includegraphics[trim = 0mm 0mm 0mm 0mm, clip, width=3.1in]{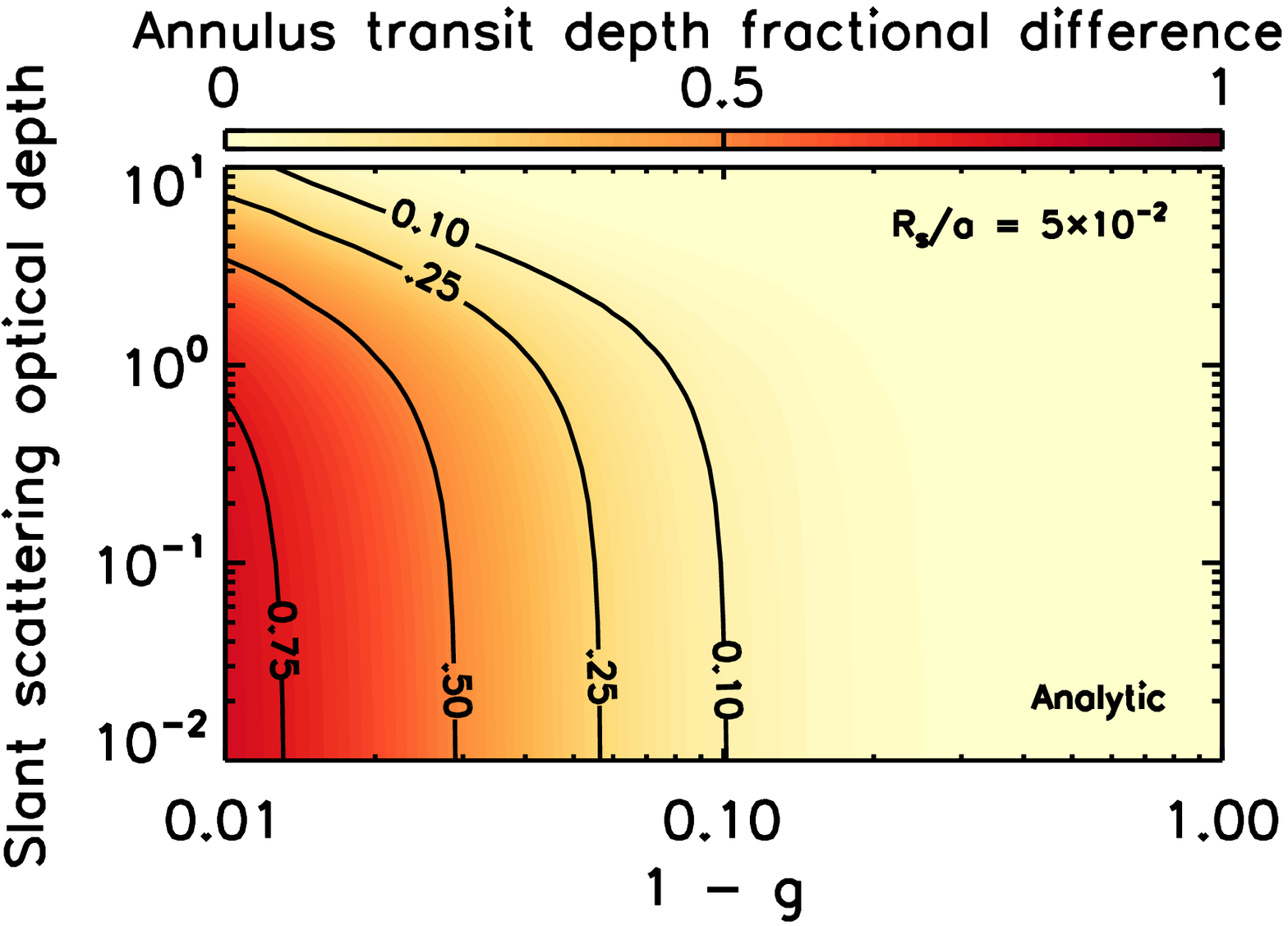} \\
    \includegraphics[trim = 0mm 0mm 0mm 0mm, clip, width=3.1in]{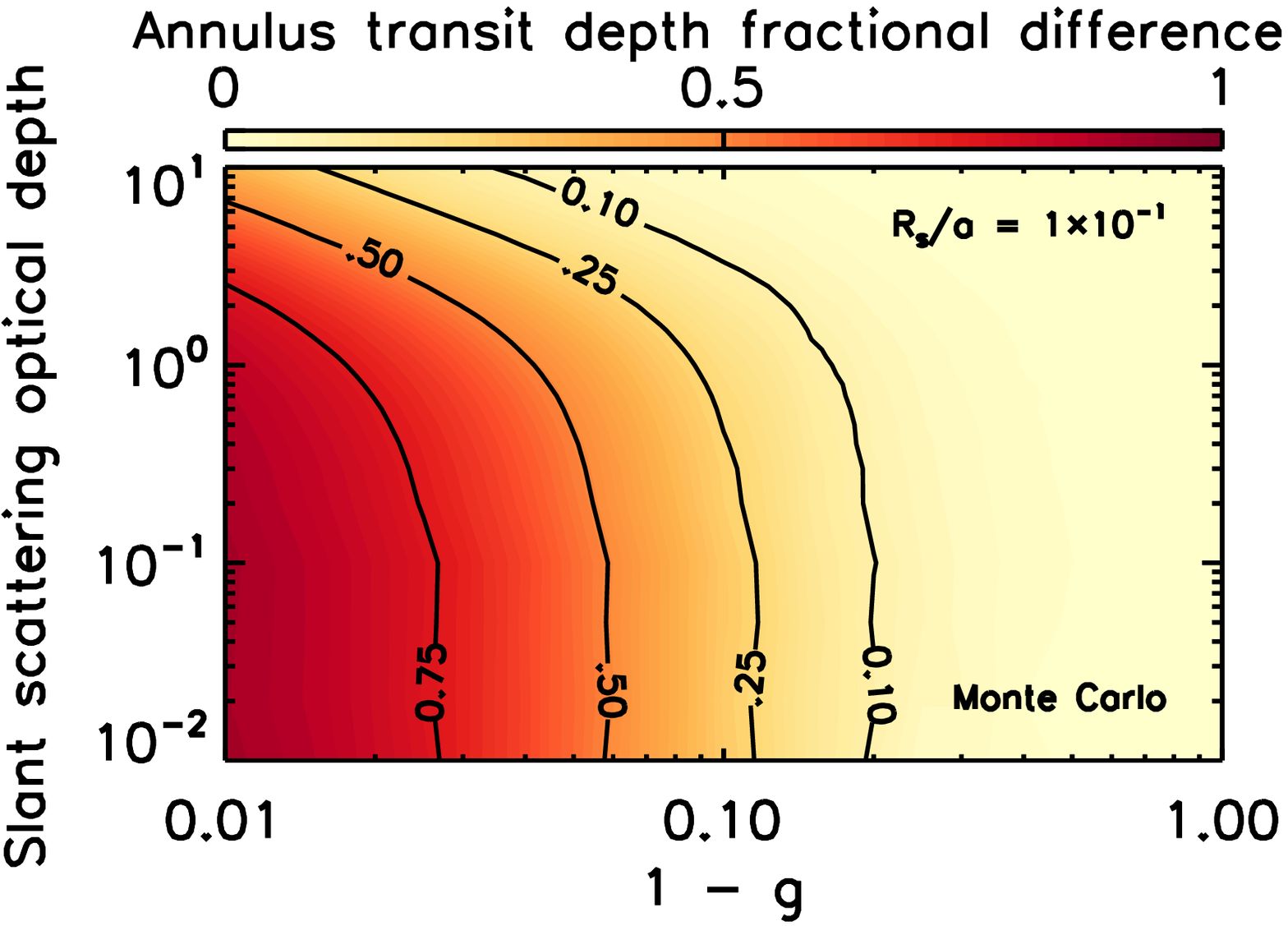}
    \includegraphics[trim = 0mm 0mm 0mm 0mm, clip, width=3.1in]{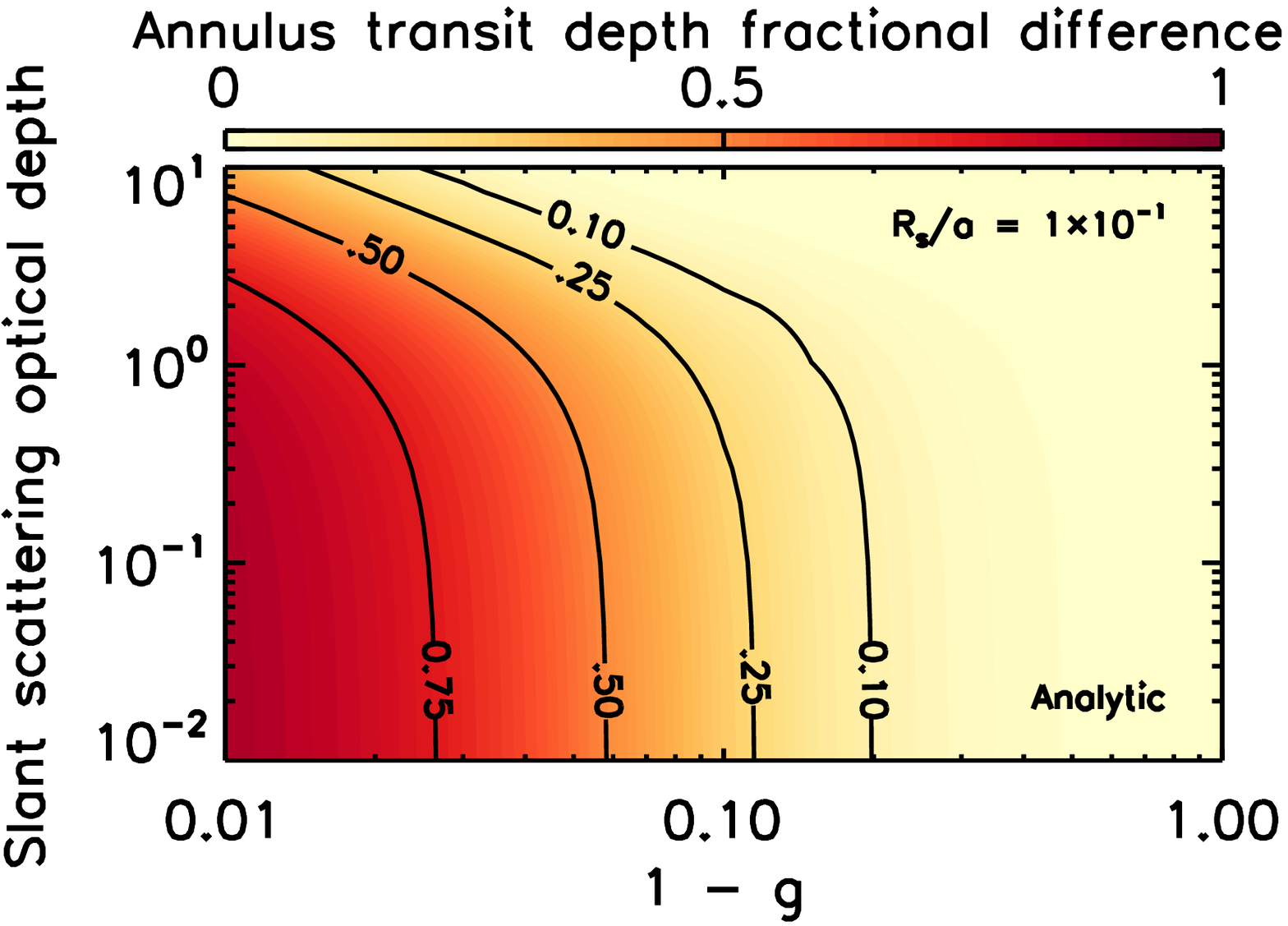} \\
  \end{tabular}
  \caption{Relative difference between the transit depth due to a single annulus in the 
                pure absorption limit versus a models that include scattering effects.  The left column 
                shows this comparison for the Monte Carlo models of \citet{robinson2017}, while the 
                right column shows this comparison when our analytic correction is used.  Sub-figures are 
                for different angular sizes of the host star as seen from the planet, and results are 
                given as a function of scattering slant optical depth within the annulus and the 
                scattering asymmetry parameter.}
  \label{fig:comp}
\end{figure}

{ We verify our scattering correction by, first, reproducing a numerical experiment 
from \citet[][their Section~4.2]{robinson2017}.  Here, the transit depth of a narrow, cloud-filled 
annulus on the planetary disk was computed for a range of cloud slant scattering optical 
depths, scattering asymmetry parameters, and host star angular sizes.  Beyond the cloud, 
no other opacity source was included, and the cloud was taken to be pure scattering with 
a single scattering phase function described by the Henyey-Greenstein phase function.  Results 
were referenced to the ``pure absorption'' scenario, where all scattering optical depth is 
treated as absorption optical depth.  Figure~\ref{fig:comp} shows the Monte Carlo cases in the 
left column and the cases where the analytic correction is used in the right column.  The transit 
depth in the analytic case simply scales as $1-{\rm{EXP}}\left[(1-f)\Delta\tau_{\rm{s}}\right]$, 
where $\Delta\tau_{\rm{s}}$ is the slant scattering optical depth of the cloud.  Clearly the 
analytic model reproduces the shape and scale of the results from the full Monte Carlo 
simulation.}

\begin{figure}
  \centering
  \includegraphics[trim = 0mm 0mm 0mm 0mm, clip, width=6.2in]{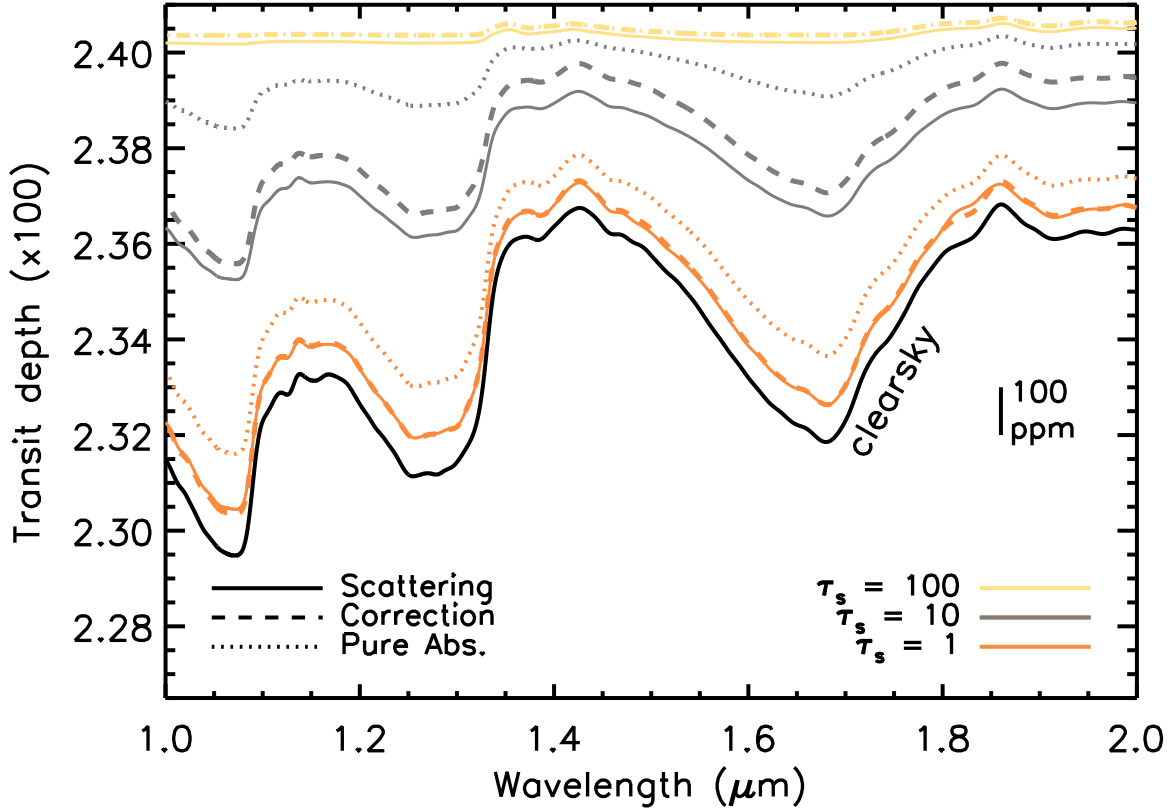}
  \caption{Transit spectra from the fully multiple scattering model of \citet[][solid]{robinson2017}, a 
                model in the pure absorption limit (dotted) where all optical depth is treated as absorption 
                optical depth, and from a model that adopts the correction factor in Equation~\ref{eqn:fHG} 
                (dashed).  Isolated, conservatively scattering clouds with different slant scattering optical depths 
                (indicated by color) are distributed across one pressure scale height centered at $10^{-4}$~bar 
                in a hot Jupiter-like atmosphere.  Additional model details are given in the text.  An asymmetry 
                parameter of $g=0.95$ is adopted.  Features are due to water vapor, and a 100 ppm bar is given 
                for scale.}
  \label{fig:scatspec}
\end{figure}

{Next we verify our scattering correction by comparing to full transit spectra computed 
according to the full-physics Monte Carlo model of \citet{robinson2017}.  For this comparison, 
we use a standardized hot Jupiter-like atmospheric model which has a planetary radius (at the 
10 bar pressure level) of 1.16~$R_{\rm{J}}$, a planetary mass of 1.14~$M_{\rm{J}}$, a stellar 
radius of 0.78~$R_{\odot}$, and atmospheric volume mixing ratios of H$_2$, He, and H$_2$O 
of 0.85, 0.15, and $4\times10^{-4}$, respectively.  The 126 model layers are placed evenly in 
log-pressure between 10 bar and 10$^-9$ bar.  The atmosphere is isothermal at 1500~K and 
the planet is located 0.031~au from its host.  Non-absorbing, forward scattering ($g=0.95$) 
clouds of different slant scattering optical depths are distributed across a pressure scale 
height, and are placed at different atmospheric pressure levels.  Figure~\ref{fig:scatspec} 
compares transit spectra that use our proposed correction to models in the pure absorption 
limit as well as to full-physics multiple scattering models.  Briefly, transit spectra using our 
correction were computed using Equations (5), (7), and (36) from \citet{robinson2017}, where 
the differential optical depth in these expressions was replaced with our effective optical 
depth.  Clearly, the model that adopts our correction performs much better than the 
commonly-used pure absorption model.  Differences between the former and the multiple 
scattering model are typically less than 30--50 ppm, while the latter typically has errors larger 
than 100--200 ppm (for the two thinnest cloud cases).}

{ While expressions similar to Equation~\ref{eqn:pmax} have previously appeared in 
the literature \citep{sidis&sari2010,betremieux&kaltenegger2014}, little work has been done to 
explore the validity of this treatment.}  We verify Equation~\ref{eqn:pmax} { (including the planetary 
radius term on the right hand side)} by comparing to the numerically derived values from 
\citet{betremieux2016}, who investigated the location of refractive boundaries in transiting 
Jupiter-like exoplanets.  Following this work, we adopt $p_0$ to be 1~bar, $\nu_0$ to have a 
Jupiter-like value of $1.23\times10^{-4}$ at STP, and $r_0$ to be equal to Jupiter's radius of 
$6.99\times10^4$~km.  Additionally, like \citet[][their Equation~2]{betremieux2016}, we set the 
atmospheric temperature, $T$, to be equal to the equilibrium temperature for a world with Bond 
albedo, $A_{\rm{B}}$, and negligible internal heat, so that, 
\begin{equation}
  \frac{a}{R_{\rm{s}}} = \frac{\sqrt{1 - A_{\rm{B}}}}{2} \left( \frac{T_{\rm{s}}}{T} \right)^{\!\!2} \ ,
\label{eqn:energybal}
\end{equation}
where $T_{\rm{s}}$ is the host stellar effective temperature, and the Bond albedo is taken 
to be 0.3.  { Table~3 compares the maximum pressures probed from Equation~\ref{eqn:pmax} 
to the numerically-derived values from \citet{betremieux2016} for the planetary temperatures 
and host star temperatures used in that study.  Here we see that the analytic expression 
reproduces the ray-tracing results to within 20\% when $p_{\rm{max}}$ is below 10~bar, 
growing to many tens of percent when $p_{\rm{max}}$ is above 10~bar.  For the cases with 
the larger values of $p_{\rm{max}}$, the angular size of the host star ($R_{\rm{s}}/a$), and 
thus the refraction angle, is greater than about 0.03,} so our assumption that rays pass straight 
through the atmosphere breaks down---these rays are, instead, bent more significantly 
towards lower pressures and probe deeper regions of the atmosphere than our simple theory 
would indicate.  However, this regime is unlikely to be important because molecular or cloud 
opacity will likely prevent rays from probing this deep into an atmosphere.

We can further test the simple treatment of refraction against a ray tracing model 
described in \citet{robinson2017}.  For this test, we consider a planet with Earth's radius 
and mass orbiting at 1~au from a Sun-like star.  We adopt an isothermal atmosphere at 
255~K (i.e., Earth's equilibrium temperature), and an atmospheric composition of N$_2$ and 
H$_2$O with volume mixing ratios of 0.99 and 0.01, respectively.  According to 
Equation~\ref{eqn:pmax}, this world would have a $p_{\rm{max}}$ of 0.18~bar at 1~$\mu$m, 
which occurs at an altitude of 12~km.  { This value is in good agreement with results from 
\citet{garciamunozetal2012}, who found a refractive floor in Earth's transit spectrum at 
11.6~km using analytic arguments and 13.2~km using numerical models.  The analytic 
result from these authors assumed an isothermal atmosphere with a pressure scale 
height of 8~km, which is nearly identical to the scale height in our isothermal case.  Using 
numerical models, \citet{misraetal2014} found a $p_{\rm{max}}$ of 0.3~bar for a realistic 
Earth/Sun pair and \citet{betremieux&kaltenegger2014} found a $p_{\rm{max}}$ of 0.17~bar.  
Differences amongst these predicted values for the location of a refractive boundary are likely 
due to the treatment of atmospheric temperature (i.e., isothermal for analytic cases versus 
varying with altitude in the numerical cases) and, as mentioned in \citet{robinson2017}, the 
integration lengths used within numerical ray tracing schemes.}

\begin{figure}
  \centering
  \begin{tabular}{c}
    \includegraphics[trim = 0mm 0mm 0mm 0mm, clip, width=3.1in]{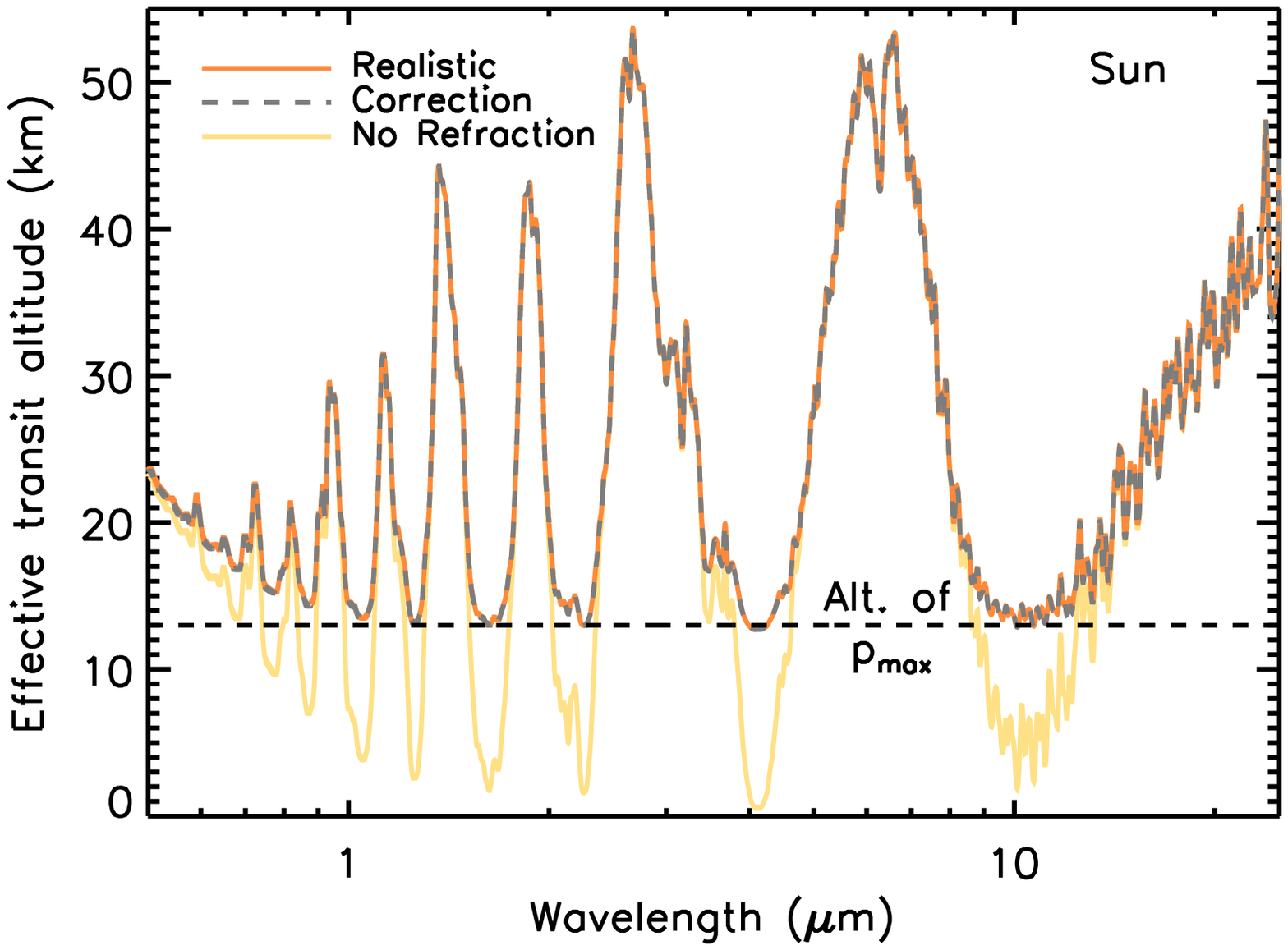} \\
    \includegraphics[trim = 0mm 0mm 0mm 0mm, clip, width=3.1in]{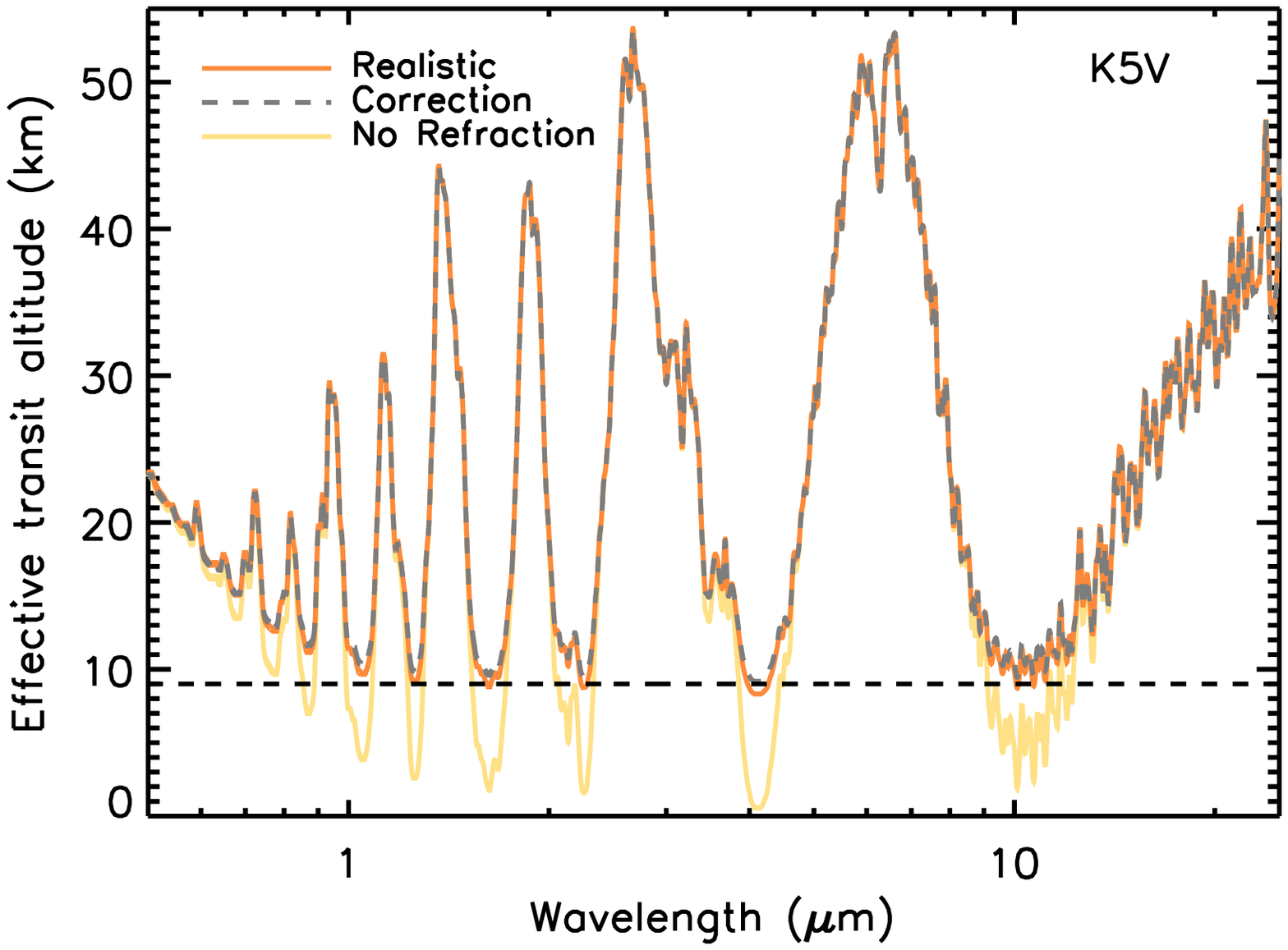} \\
    \includegraphics[trim = 0mm 0mm 0mm 0mm, clip, width=3.1in]{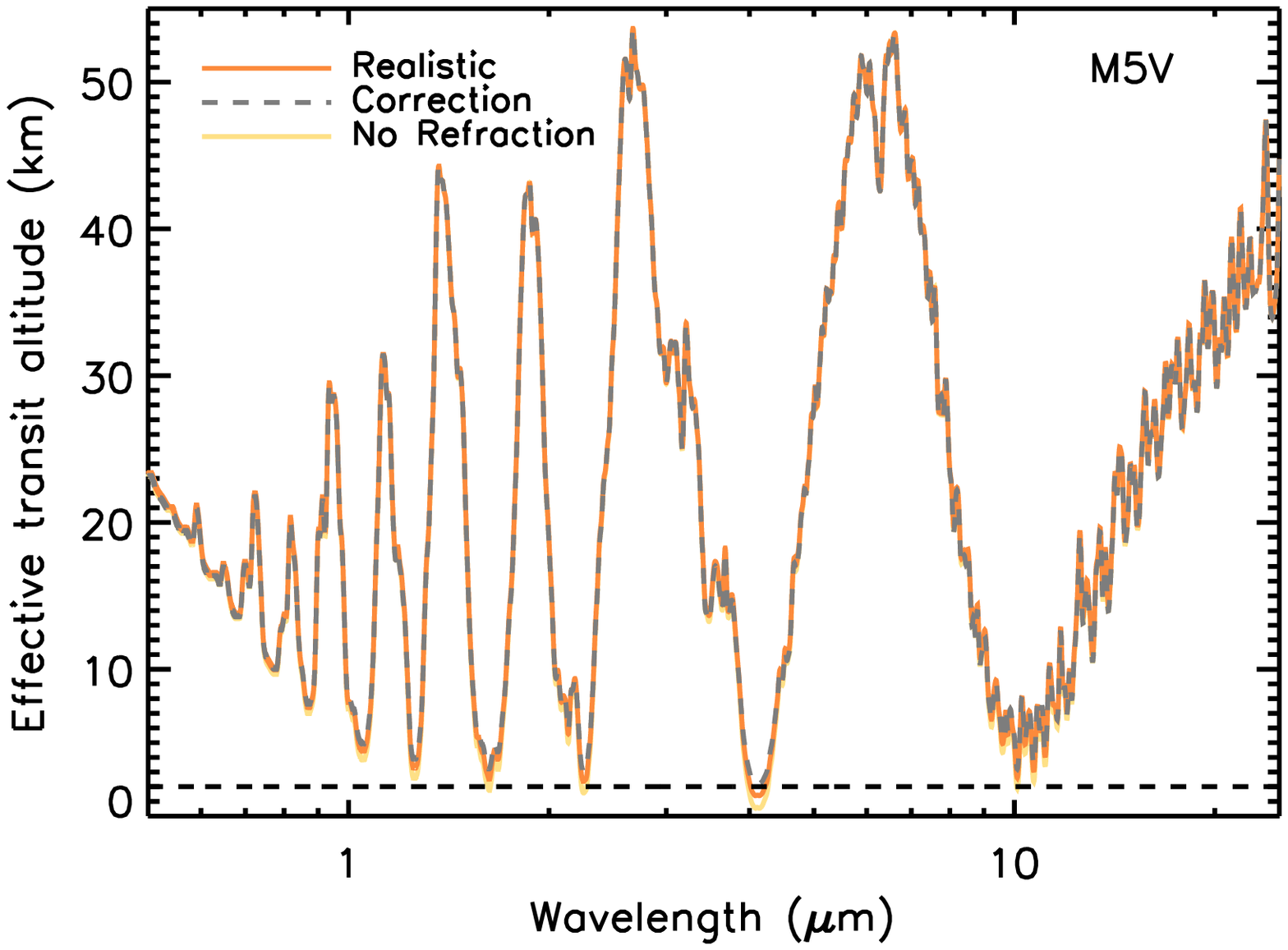} \\
  \end{tabular}
  \caption{Transit spectra of an Earth-sized exoplanet at the Earth equivalent insolation distance 
               for a Sun-like (top), K5 dwarf (middle), and M5 dwarf (bottom) host.  Model atmosphere 
               details are provided in the text.  For each case, the refractive boundary from 
               Equation~\ref{eqn:pmax} is indicated by a dashed black line.  Colors indicate spectra 
               from a ray tracing model (orange), a model without refraction (yellow), 
               and a model that includes a simple correction for refraction (gray) where the transmission 
               is set to zero for rays that probe deeper than $p_{\rm{max}}$.}
  \label{fig:simple}
\end{figure}

When computing a transit spectrum by performing an integral/sum over a grid of impact 
parameters \citep[e.g.,][their Equation~4]{betremieux&kaltenegger2013}, one 
simply sets the transmission to zero for all rays with impact parameters below this refractive 
boundary.  Figure~\ref{fig:simple} shows transit spectra for our 
adopted Earth-sized planet produced by a fully realistic model with refraction, a model without 
refraction, and a non-refracting model that sets slant transmission to zero below the refractive 
boundary at 12~km above the surface.  (Note that the effective transit altitude, $z_{\rm{eff}}$, is 
defined by setting the wavelength-dependent transit depth equal to 
$\left[\left(R_{\rm{p}} + z_{\rm{eff}}\right)/R_{\rm{s}}\right]^2$.) The model with the analytic 
refractive boundary improves quite significantly over a model that does not include refraction. 



Expanding our calculation to cooler spectral types, we find that the refractive boundary 
for dwarf K5, M0, and M5 stars occurs at 32~mbar (9~km), 41~mbar (7~km), and 
78~mbar (2~km), respectively, where we place the planet at the Earth equivalent 
insolation distance (0.37, 0.27, and 0.047~au, respectively).  { These values are in 
general agreement with the ray-tracing results of \citet{betremieux&kaltenegger2014} 
for Earths around host stars of different spectral type, and small differences (less than 
2~km) can likely be attributed to these authors using a realistic Earth temperature-pressure 
profile.}  Figure~\ref{fig:simple} shows transit spectra of our Earth-sized planet for the 
K5 and M5 dwarf cases.  For all cases, our approach of setting the transmission to zero 
for rays that probe deeper than $p_{\rm{max}}$ reproduces the ray tracing model quite 
well.


%
\subsection{Refractive Boundaries in Jovian Atmospheres with Collision Induced Absorption}

An open question in exoplanet transit spectroscopy and spectral retrieval is the degree 
to which refraction may influence transit observations of worlds with hydrogen rich 
atmospheres.  Refraction can be of first-order importance for worlds with 
nitrogen rich atmospheres \citep{betremieux&kaltenegger2014,misraetal2014}, but this 
is due to molecular nitrogen's lack of broad absorption bands or collision induced absorption 
(CIA) features \citep[except near 4.5~$\mu$m][]{schietermanetal2015}.  Molecular hydrogen 
has a broad CIA continuum which will compete with refraction \citep[as well as clouds and 
Rayleigh scattering,][]{sidis&sari2010} to set the limiting pressures probed between molecular 
absorption bands in transit spectrum observations.  Here we extend the study of 
\citet{betremieux2016}, who investigated refraction effects in gas giant atmospheres but omitted 
both H$_2$-H$_2$ and H$_2$-He CIA.

\begin{figure}
  \centering
  \begin{tabular}{c}
    \includegraphics[trim = 0mm 0mm 0mm 0mm, clip, width=3.1in]{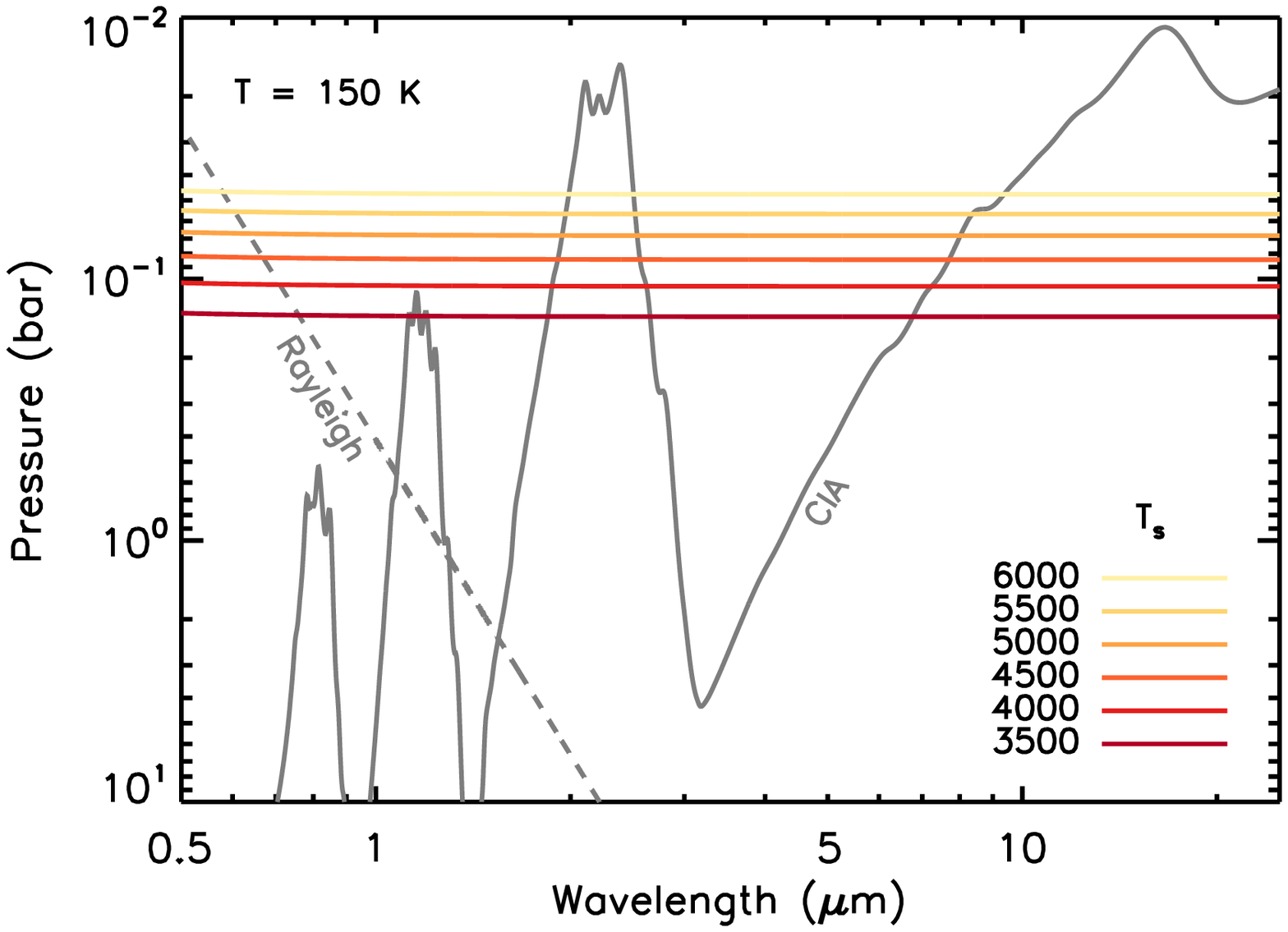} \\
    \includegraphics[trim = 0mm 0mm 0mm 0mm, clip, width=3.1in]{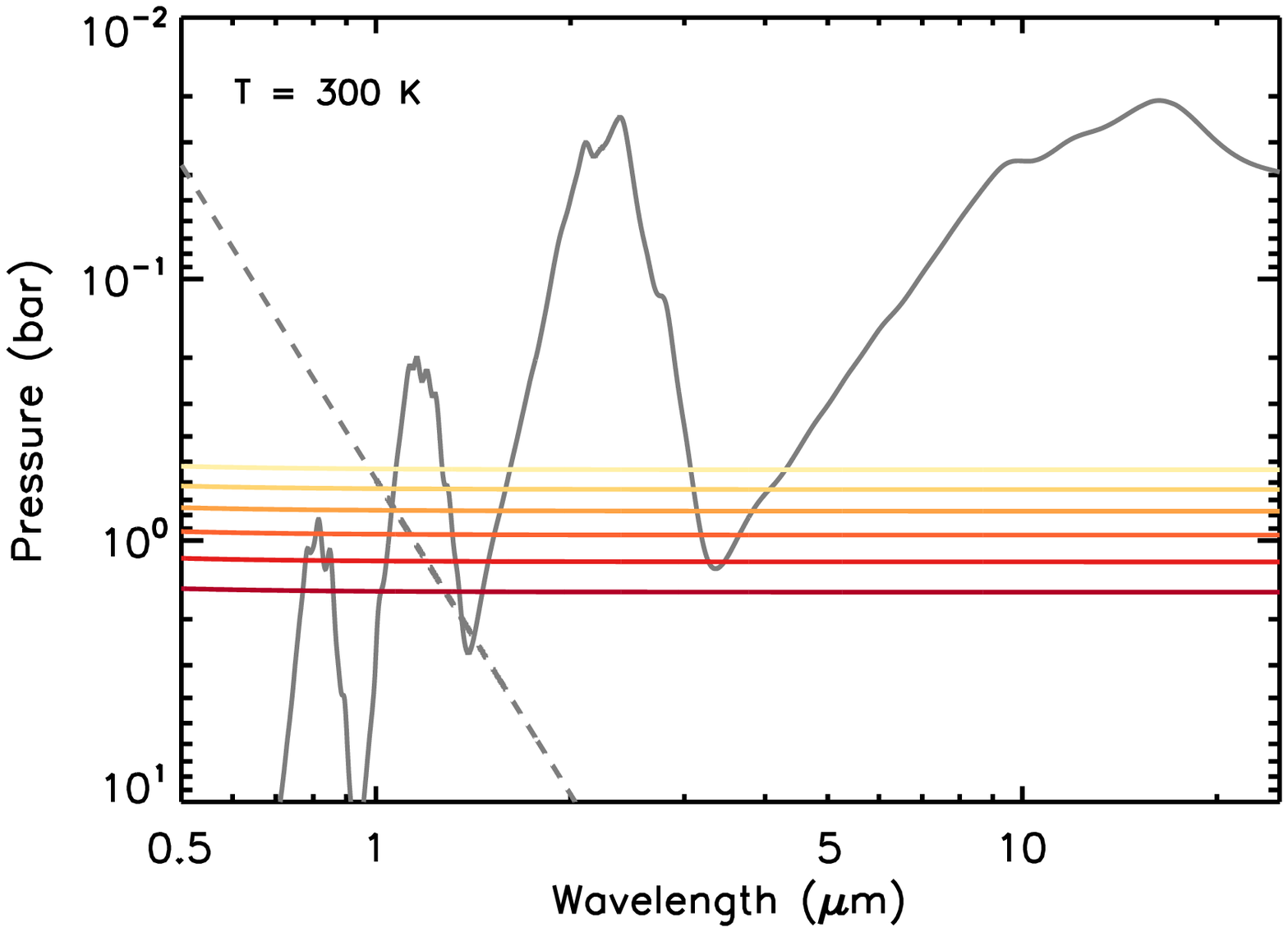} \\
    \includegraphics[trim = 0mm 0mm 0mm 0mm, clip, width=3.1in]{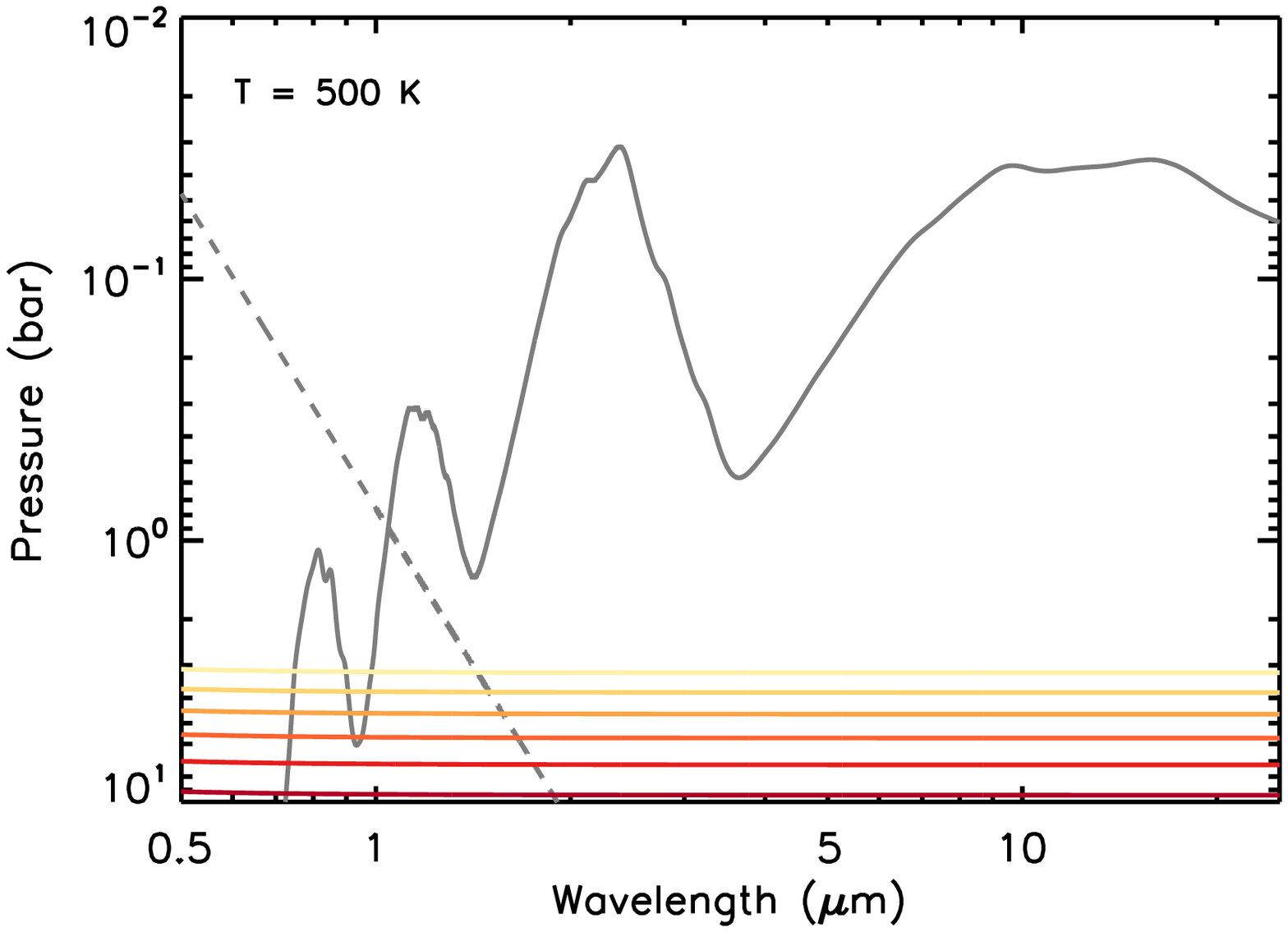} \\
  \end{tabular}
  \caption{Pressure of slant optical depth equal to 0.5 for Rayleigh scattering (gray dashed) and 
                CIA (gray solid) for Jovian worlds of different temperatures.  Colored lines indicate the 
                pressure of the refractive boundary (from Equation~\ref{eqn:pmax}) for host stars of 
                different effective temperatures, where the angular size of the star is set by the energy 
                balance required to yield the planetary temperature (Equation~\ref{eqn:energybal}).  When 
                horizontal colored lines fall below the gray curves, the refractive boundary is located at 
                a higher pressure (or smaller altitude/radius) than where the transit spectrum would become  
                opaque to Rayleigh scattering or CIA opacity, implying that refraction would not influence the 
                shape of the transit spectrum.}
  \label{fig:ciarefcomp}
\end{figure}

Figure~\ref{fig:ciarefcomp} shows the pressure where the slant optical depth equals 
0.56 for Rayleigh scattering and CIA for Jupiter-like worlds with isothermal atmospheres 
at 150, 300, and 500~K.  We assume a radius of $6.99\times10^4$~km, a surface gravity 
of 24.8~m~s$^{-2}$, and a 0.9/0.1 mixture of H$_2$/He (by number).  Rayleigh scattering 
and CIA from both hydrogen and helium \citep{abeletal2011} are included.  Also shown are 
the pressures of the refractive boundary for host stars of different effective temperatures, from 
{ the non-approximate} Equation~\ref{eqn:pmax}.  The angular size of the host star is 
determined using the energy balance expression in Equation~\ref{eqn:energybal} for an 
assumed Bond albedo of 0.3.  The pressure of the refractive boundary will depend weakly on 
Bond albedo, scaling as $\left(1-A_{\rm{B}}\right)^{-1/2}$.  

In Figure~\ref{fig:ciarefcomp}, wavelengths where the refractive boundary curves sit 
above the Rayleigh or CIA curves are regions where refraction has the potential to limit 
observations (in the absence of any molecular absorption features from trace atmospheric 
species).  The highest equilibrium temperatures where refraction will have a significant effect 
is roughly 400--500~K, depending on wavelength.  For higher temperatures, the host star 
(as seen from the planet) has a relatively large angular size, thereby pushing the refractive 
boundary to pressures deeper than about 1--3 bar. 

\section{Discussion}

The expressions given above are designed to be easily incorporated into existing transit 
spectrum models with little additional computational expense.  For aerosol forward 
scattering, one only need evaluate the correction factor (e.g.,~Equation~\ref{eqn:fHG}) 
for each layer while summing optical depths along a slant path.  As shown in 
Figure~\ref{fig:scatspec}, this correction makes a clear improvement over the 
commonly-used pure absorption models.  As outlined in Section~\ref{subsec:scat}, 
our forward scattering correction can be used in cases where the angular symmetry 
about the planet disk is broken, and would result in more muted effects from 
asymmetric forward scattering clouds on exoplanet terminators than are currently 
assumed in simulations \citep{line&parmentier2016,macdonald&madhusudhan2017}.

Our analysis, and as shown in Figure~\ref{fig:dzH}, indicates that omitting a forward 
scattering correction can cause a transit spectrum to be biased towards higher 
altitudes by as many as four aerosol scale heights.  In other words, the effective 
transit altitude could be too large by several aerosol scale heights.  Of course, this 
biasing depends on the host star angular size and the scattering phase function, 
amongst other parameters.  As transit spectra typically probe several {\it pressure} 
scale heights, Figure~\ref{fig:dzH} shows that forward scattering effects can become 
comparable to molecular features when the aerosol scale height is comparable to 
(or larger than) the pressure scale height, when $g \gtrsim 0.85$, and when 
$R_{\rm{s}}/a \gtrsim 0.1$.

Regarding refraction, Equation~\ref{eqn:pmax} { (see also \citet{sidis&sari2010}, 
their Equation~17, and \citet{betremieux&kaltenegger2014}, their Equation~11)} 
allows modelers and observers to quickly evaluate whether or not a refractive boundary 
is likely to influence a given transit spectrum.  { Comparisons to previously-published 
results from a ray-tracing model applied to gas giant exoplanets  \citep{betremieux2016} 
indicates that the expression is valid to within 10--20\% for angular size of the host star 
($R_{\rm{s}}/a$) below about 0.03 (which correspond to maximum pressures below 
roughly 10~bar; see Table~2).}  At larger pressures, our assumption 
of a straight-line path for our rays breaks down, as rays would actually be bent to probe 
deeper into the exoplanet atmosphere.  However, gas and aerosol extinction are likely to 
prevent observations from reaching such large pressures---the typical pressures probed 
in transit spectra are fractions of a bar.

In practice, if Equation~\ref{eqn:pmax} indicates that refraction may be an important 
consideration for a certain model or observation, then the effects of a refractive boundary 
can be mimicked by simply setting the slant transmission to zero below the refractive 
boundary when integrating over concentric annuli to produce a transit spectrum.  
Figure~\ref{fig:simple} demonstrates the efficacy of this simple approach across a wide 
range of angular sizes of the stellar host for Earth-sized planets with cloud-free, 1~bar 
N$_{2}$-H$_2$O atmospheres.

For retrieval studies, our forward scattering correction factor shows that any cloud 
optical depth retrieved from an exoplanet transit observation is actually an ``effective'' 
optical depth that folds together information about the extent of the forward scattering 
peak in the scattering phase function and the angular size of the host star.  Also, 
Equation~\ref{eqn:pmax} indicates that, should a refractive boundary be detected in 
an exoplanet transit spectrum, inverse analyses may be able to use the location of this 
boundary to help constrain background atmospheric constituents, as this expression 
depends on the bulk atmospheric refractivity.  {However, the transit depth down to 
this refractive boundary depends on both the atmospheric scale height and the bulk 
refractivity \citep{sidis&sari2010}.  Independently measuring the atmospheric scale 
height (through, e.g., detecting a Rayleigh scattering slope 
\citep{lecavelierdesetangsetal2008}) would break this degeneracy.  This characteristic 
of refractive boundaries is in addition to their ability to help break degeneracies when 
attempting to constrain abundances of trace atmospheric constituents 
\citep{betremieux2016,betremieux&swain2017}.}

Finally, our analysis of the competing effects of refractive boundaries and Rayleigh 
scattering and CIA for Jupiter-like exoplanets provides a clear indication of the regimes 
where refraction will outweigh opacity from H$_2$ and He.  As shown in 
Figure~\ref{fig:ciarefcomp}, refraction can outweigh Rayleigh scattering and CIA for 
worlds whose equilibrium temperatures are below about 400--500~K.  For equilibrium 
temperatures above this, the planet is located relatively close to its host star, implying 
a large host star angular size which pushes the refractive boundary to pressures 
deeper than about 1--3 bar.  Note that, for the case in Figure~\ref{fig:ciarefcomp} with 
an atmospheric temperature of 500~K, the planets are located at orbital distances 
of 0.05--0.3~au from the hosts with effective temperatures of 3500--6000~K, and 
have orbital periods of 0.01--0.2~yr.

\section{Conclusions}

Refraction and aerosol forward scattering are complicated physical processes and, 
as a result, are difficult to efficiently incorporate into standard transit spectrum tools.  
We have derived analytic expressions that can be used to easily account for 
scattering and a refractive boundary in transit observations and models.  For scattering, 
we derive an effective slant optical depth that includes a correction for light that is 
contained in the forward scattering peak of a aerosol scattering phase function, and 
we demonstrate the accuracy of this correction by comparing to full-physics models.  
Exploring this correction shows that transit spectra for exoplanets with 
exponentially-distributed hazes can be biased by up to four scale heights when forward 
scattering is ignored, depending on the angular size of the host star and the scattering 
asymmetry parameter.  { We validate the utility of a common expression for 
the location of a refractive boundary, which we write in terms of the refractive boundary 
pressure for easy incorporation into existing transit spectrum models.}  Using { the 
validated expression} for refractive boundaries, we show that opacity from Rayleigh 
scattering and CIA will outweigh refraction effects for Jupiter-like worlds with equilibrium 
temperatures above 400--500~K.

\acknowledgements
TR gratefully acknowledges support from NASA through the Sagan Fellowship 
Program executed by the NASA Exoplanet Science Institute. The results reported 
herein benefitted from collaborations and/or information exchange within NASA's 
Nexus for Exoplanet System Science (NExSS) research coordination network 
sponsored by NASA's Science Mission Directorate.  Certain essential tools used 
in this work were developed by the NASA Astrobiology Institute's Virtual Planetary 
Laboratory, supported by NASA under Cooperative Agreement No. NNA13AA93A.


\begin{thebibliography}{}
\expandafter\ifx\csname natexlab\endcsname\relax\def\natexlab#1{#1}\fi

\bibitem[{{Abel} {et~al.}(2011){Abel}, {Frommhold}, {Li}, \&
  {Hunt}}]{abeletal2011}
{Abel}, M., {Frommhold}, L., {Li}, X., \& {Hunt}, K.~L.~C. 2011, Journal of
  Physical Chemistry A, 115, 6805

\bibitem[{{Barstow} {et~al.}(2012){Barstow}, {Aigrain}, {Irwin}, {Bowles},
  {Fletcher}, \& {Lee}}]{barstowetal2012}
{Barstow}, J.~K., {Aigrain}, S., {Irwin}, P.~G.~J., {et~al.} 2012, ArXiv
  e-prints, arXiv:1212.5020
  
  \bibitem[Baum \& Code(1953)]{baum&code1953} Baum, W.~A., \& Code, A.~D.\ 1953, \aj, 58, 108 

\bibitem[{Beichman {et~al.}(2014)Beichman, Benneke, Knutson, Smith, Lagage,
  Dressing, Latham, Lunine, Birkmann, Ferruit, {et~al.}}]{beichmanetal2014}
Beichman, C., Benneke, B., Knutson, H., {et~al.} 2014, Publications of the
  Astronomical Society of the Pacific, 126, 1134

\bibitem[{Benneke \& Seager(2012)}]{benneke&seager2012}
Benneke, B., \& Seager, S. 2012, The Astrophysical Journal, 753, 100

\bibitem[{{B{\'e}tr{\'e}mieux}(2016)}]{betremieux2016}
{B{\'e}tr{\'e}mieux}, Y. 2016, \mnras, 456, 4051

\bibitem[{B{\'e}tr{\'e}mieux \& Kaltenegger(2013)}]{betremieux&kaltenegger2013}
B{\'e}tr{\'e}mieux, Y., \& Kaltenegger, L. 2013, The Astrophysical Journal
  Letters, 772, L31

\bibitem[{{B{\'e}tr{\'e}mieux} \&
  {Kaltenegger}(2014)}]{betremieux&kaltenegger2014}
{B{\'e}tr{\'e}mieux}, Y., \& {Kaltenegger}, L. 2014, \apj, 791, 7

\bibitem[{B{\'e}tr{\'e}mieux \& Kaltenegger(2015)}]{betremieux&kaltenegger2015}
B{\'e}tr{\'e}mieux, Y., \& Kaltenegger, L. 2015, Monthly Notices of the Royal
  Astronomical Society, 451, 1268

\bibitem[{{B{\'e}tr{\'e}mieux} \& {Swain}(2016)}]{betremieux&swain2017}
{B{\'e}tr{\'e}mieux}, Y., \& {Swain}, M.~R. 2016, ArXiv e-prints,
  arXiv:1610.02049

\bibitem[{{Bideau-Mehu} {et~al.}(1973){Bideau-Mehu}, {Guern}, {Abjean}, \&
  {Johannin-Gilles}}]{bideaumehuetal1973}
{Bideau-Mehu}, A., {Guern}, Y., {Abjean}, R., \& {Johannin-Gilles}, A. 1973,
  Optics Communications, 9, 432

\bibitem[{Brown(2001)}]{brown2001}
Brown, T.~M. 2001, The Astrophysical Journal, 553, 1006

\bibitem[{{Dalba} {et~al.}(2015){Dalba}, {Muirhead}, {Fortney}, {Hedman},
  {Nicholson}, \& {Veyette}}]{dalbaetal2015}
{Dalba}, P.~A., {Muirhead}, P.~S., {Fortney}, J.~J., {et~al.} 2015, \apj, 814,
  154

\bibitem[{De~Kok \& Stam(2012)}]{dekok&stam2012}
De~Kok, R., \& Stam, D. 2012, Icarus, 221, 517

\bibitem[Fabry(1929)]{fabry1929} Fabry, C.\ 1929, Journal des Observateurs, 12, 1 

\bibitem[{{Fraine} {et~al.}(2014){Fraine}, {Deming}, {Benneke}, {Knutson},
  {Jord{\'a}n}, {Espinoza}, {Madhusudhan}, {Wilkins}, \&
  {Todorov}}]{fraineetal2014}
{Fraine}, J., {Deming}, D., {Benneke}, B., {et~al.} 2014, \nat, 513, 526

\bibitem[Garc{\'{\i}}a Mu{\~n}oz et al.(2012)]{garciamunozetal2012} Garc{\'{\i}}a Mu{\~n}oz, 
A., Zapatero Osorio, M.~R., Barrena, R., et al.\ 2012, \apj, 755, 103 

\bibitem[{{Gardner} {et~al.}(2006){Gardner}, {Mather}, {Clampin}, {Doyon},
  {Greenhouse}, {Hammel}, {Hutchings}, {Jakobsen}, {Lilly}, {Long}, {Lunine},
  {McCaughrean}, {Mountain}, {Nella}, {Rieke}, {Rieke}, {Rix}, {Smith},
  {Sonneborn}, {Stiavelli}, {Stockman}, {Windhorst}, \&
  {Wright}}]{gardneretal2006}
{Gardner}, J.~P., {Mather}, J.~C., {Clampin}, M., {et~al.} 2006, \ssr, 123, 485

\bibitem[{{Goldsmith}(1963)}]{goldsmith1963}
{Goldsmith}, D.~W. 1963, Icarus, 2, 341

\bibitem[{{Henyey} \& {Greenstein}(1941)}]{henyey&greenstein1941}
{Henyey}, L.~G., \& {Greenstein}, J.~L. 1941, \apj, 93, 70

\bibitem[{{Hill} \& {Lawrence}(1986)}]{hill&lawrence1986}
{Hill}, R.~J., \& {Lawrence}, R.~S. 1986, Infrared Physics, 26, 371

\bibitem[{Howe \& Burrows(2012)}]{howe&burrows2012}
Howe, A.~R., \& Burrows, A.~S. 2012, The Astrophysical Journal, 756, 176

\bibitem[{Hubbard {et~al.}(2001)Hubbard, Fortney, Lunine, Burrows, Sudarsky, \&
  Pinto}]{hubbardetal2001}
Hubbard, W., Fortney, J., Lunine, J., {et~al.} 2001, The Astrophysical Journal,
  560, 413

\bibitem[{Hubbard {et~al.}(1988)Hubbard, Hunten, Dieters, Hill, \&
  Watson}]{hubbardetal1988}
Hubbard, W., Hunten, D., Dieters, S., Hill, K., \& Watson, R. 1988, Nature,
  336, 452

\bibitem[{{Irvine}(1965)}]{irvine1965}
{Irvine}, W.~M. 1965, \apj, 142, 1563

\bibitem[{{Johnson} {et~al.}(2010){Johnson}, {Aller}, {Howard}, \&
  {Crepp}}]{johnsonetal2010}
{Johnson}, J.~A., {Aller}, K.~M., {Howard}, A.~W., \& {Crepp}, J.~R. 2010,
  \pasp, 122, 905

\bibitem[{{Knutson} {et~al.}(2014){Knutson}, {Dragomir}, {Kreidberg},
  {Kempton}, {McCullough}, {Fortney}, {Bean}, {Gillon}, {Homeier}, \&
  {Howard}}]{knutsonetal2014b}
{Knutson}, H.~A., {Dragomir}, D., {Kreidberg}, L., {et~al.} 2014, \apj, 794,
  155

\bibitem[{{Kreidberg}(2015)}]{kreidberg2015}
{Kreidberg}, L. 2015, \pasp, 127, 1161

\bibitem[{{Kreidberg} {et~al.}(2014){Kreidberg}, {Bean}, {D{\'e}sert}, {Line},
  {Fortney}, {Madhusudhan}, {Stevenson}, {Showman}, {Charbonneau},
  {McCullough}, {Seager}, {Burrows}, {Henry}, {Williamson}, {Kataria}, \&
  {Homeier}}]{kreidbergetal2014b}
{Kreidberg}, L., {Bean}, J.~L., {D{\'e}sert}, J.-M., {et~al.} 2014, \apjl, 793,
  L27

\bibitem[{{K{\v r}en}(2011)}]{kren2011}
{K{\v r}en}, P. 2011, \ao, 50, 6484

\bibitem[{Lecavelier Des~Etangs {et~al.}(2008)Lecavelier Des~Etangs, Pont,
  Vidal-Madjar, \& Sing}]{lecavelierdesetangsetal2008}
Lecavelier Des~Etangs, A., Pont, F., Vidal-Madjar, A., \& Sing, D. 2008,
  Astronomy and Astrophysics, 481, L83

\bibitem[{{Lee} {et~al.}(2014){Lee}, {Irwin}, {Fletcher}, {Heng}, \&
  {Barstow}}]{leeetal2014}
{Lee}, J.-M., {Irwin}, P.~G.~J., {Fletcher}, L.~N., {Heng}, K., \& {Barstow},
  J.~K. 2014, \apj, 789, 14

\bibitem[{{Line} \& {Parmentier}(2016)}]{line&parmentier2016}
{Line}, M.~R., \& {Parmentier}, V. 2016, \apj, 820, 78

\bibitem[{{Line} {et~al.}(2012){Line}, {Zhang}, {Vasisht}, {Natraj}, {Chen}, \&
  {Yung}}]{lineetal2011}
{Line}, M.~R., {Zhang}, X., {Vasisht}, G., {et~al.} 2012, \apj, 749, 93

\bibitem[{{MacDonald} \& {Madhusudhan}(2017)}]{macdonald&madhusudhan2017}
{MacDonald}, R.~J., \& {Madhusudhan}, N. 2017, ArXiv e-prints, arXiv:1701.01113

\bibitem[{{Mandel} \& {Agol}(2002)}]{mandel&agol2002}
{Mandel}, K., \& {Agol}, E. 2002, \apjl, 580, L171

\bibitem[{{Mansfield} \& {Peck}(1969)}]{mansfield&peck1969}
{Mansfield}, C.~R., \& {Peck}, E.~R. 1969, Journal of the Optical Society of
  America (1917-1983), 59, 199

\bibitem[{{Misra} {et~al.}(2014){Misra}, {Meadows}, \& {Crisp}}]{misraetal2014}
{Misra}, A., {Meadows}, V., \& {Crisp}, D. 2014, \apj, 792, 61

\bibitem[{{Misra} \& {Meadows}(2014)}]{misra&meadows2014}
{Misra}, A.~K., \& {Meadows}, V.~S. 2014, \apjl, 795, L14

\bibitem[{{Morley} {et~al.}(2016){Morley}, {Knutson}, {Line}, {Fortney},
  {Thorngren}, {Marley}, {Teal}, \& {Lupu}}]{morleyetal2016}
{Morley}, C.~V., {Knutson}, H., {Line}, M., {et~al.} 2016, ArXiv e-prints,
  arXiv:1610.07632

\bibitem[{Mu{\~n}oz {et~al.}(2012)Mu{\~n}oz, Osorio, Barrena,
  Monta{\~n}{\'e}s-Rodr{\'\i}guez, Mart{\'\i}n, \& Pall{\'e}}]{munozetal2012}
Mu{\~n}oz, A.~G., Osorio, M.~Z., Barrena, R., {et~al.} 2012, The Astrophysical
  Journal, 755, 103

\bibitem[{{Peck} \& {Huang}(1977)}]{peck&huang1977}
{Peck}, E.~R., \& {Huang}, S. 1977, Journal of the Optical Society of America
  (1917-1983), 67, 1550

\bibitem[{{Peck} \& {Nath Khanna}(1966)}]{peck&nathkhanna1966}
{Peck}, E.~R., \& {Nath Khanna}, B. 1966, Journal of the Optical Society of
  America (1917-1983), 56, 1059

\bibitem[{Petty(2006)}]{petty2006}
Petty, G. 2006, A First Course in Atmospheric Radiation (Sundog Publishing)

\bibitem[{Polyanskiy(2016)}]{mikhail_refract}
Polyanskiy, M.~N. 2016, Refractive index database,
  \url{http://refractiveindex.info}

\bibitem[{{Robinson}(2017)}]{robinson2017}
{Robinson}, T.~D. 2017, ArXiv e-prints, arXiv:1701.05564

\bibitem[{Robinson {et~al.}(2014)Robinson, Maltagliati, Marley, \&
  Fortney}]{robinsonetal2014a}
Robinson, T.~D., Maltagliati, L., Marley, M.~S., \& Fortney, J.~J. 2014,
  Proceedings of the National Academy of Sciences, 111, 9042

\bibitem[{{Schwieterman} {et~al.}(2015){Schwieterman}, {Robinson}, {Meadows},
  {Misra}, \& {Domagal-Goldman}}]{schietermanetal2015}
{Schwieterman}, E.~W., {Robinson}, T.~D., {Meadows}, V.~S., {Misra}, A., \&
  {Domagal-Goldman}, S. 2015, \apj, 810, 57

\bibitem[{Seager \& Sasselov(2000)}]{seager&sasselov2000}
Seager, S., \& Sasselov, D. 2000, The Astrophysical Journal, 537, 916

\bibitem[{{Sidis} \& {Sari}(2010)}]{sidis&sari2010}
{Sidis}, O., \& {Sari}, R. 2010, The Astrophysical Journal, 720, 904

\bibitem[{{Sing} {et~al.}(2016){Sing}, {Fortney}, {Nikolov}, {Wakeford},
  {Kataria}, {Evans}, {Aigrain}, {Ballester}, {Burrows}, {Deming},
  {D{\'e}sert}, {Gibson}, {Henry}, {Huitson}, {Knutson}, {Lecavelier Des
  Etangs}, {Pont}, {Showman}, {Vidal-Madjar}, {Williamson}, \&
  {Wilson}}]{singetal2016}
{Sing}, D.~K., {Fortney}, J.~J., {Nikolov}, N., {et~al.} 2016, \nat, 529, 59

\bibitem[{Thomas \& Stamnes(1999)}]{thomas&stamnes1999}
Thomas, G., \& Stamnes, K. 1999, Radiative Transfer in the Atmosphere and
  Ocean, Cambridge Atmospheric and Space Science Series (Cambridge University
  Press)

\bibitem[{{Waldmann} {et~al.}(2015){Waldmann}, {Tinetti}, {Rocchetto},
  {Barton}, {Yurchenko}, \& {Tennyson}}]{waldmannetal2015}
{Waldmann}, I.~P., {Tinetti}, G., {Rocchetto}, M., {et~al.} 2015, \apj, 802,
  107

\end{thebibliography}


\newpage

\section{Tables and Figures}
%

\begin{table}[ht]
  \scriptsize
  { Table 1. Gas Refractivity Expressions at STP$^{1}$} \\
  \vspace{2mm}
  \begin{tabular}{c c l l}
    \hline
    \hline
    Gas &  Refractivity, $\nu$ (with $\sigma = 1~\mu{\rm{m}}/\lambda$) & Reference & Note(s) \\
    \hline
    H$_2$    &  $1.48956\times10^{-2}/\left(180.7 - \sigma^{2}\right) + 4.9037\times10^{-3}/\left(92 - \sigma^{2}\right) $ & {\citet{peck&huang1977}} & 0.2~$\mu$m < $\lambda$ < 1.7~$\mu$m \\
      He        & $0.01470091/\left(423.98 - \sigma^{2}\right)$ & {\citet{mansfield&peck1969}}  & 0.5~$\mu$m < $\lambda$ < 2.1~$\mu$m \\
                   & $6.991\times10^{-2}/\left(166.175 - \sigma^{2}\right)$ &     &   \\
                    & $+ 1.4472\times10^{-3}/\left(79.609 - \sigma^{2}\right) $ &     &   \\
  CO$_2$   & $+ 6.42941\times10^{-5}/\left(56.3064 - \sigma^{2}\right) $ & {\citet{bideaumehuetal1973}} &   0.2~$\mu$m < $\lambda$ < 2.5~$\mu$m  \\
                  & $+ 5.21306\times10^{-5}/\left(46.0196 - \sigma^{2}\right)  $ & &  \\
                  & $+ 1.46847\times10^{-6}/\left(0.0584738 - \sigma^{2}\right)  $ & &  \\ 
    N$_2$   & $6.8552\times10^{-5} + 3.243157\times10^{-2}/\left(144 - \sigma^{2}\right)$ & {\citet{peck&nathkhanna1966}}  & 0.2~$\mu$m < $\lambda$ < 2.5~$\mu$m \\
    O$_2$   & $1.26805\times10^{-4} + 1.04202\times10^{-2}/\left(75.4 - \sigma^{2}\right)$ & {\citet{kren2011}}  & 0.4~$\mu$m < $\lambda$ < 1.8~$\mu$m \\
                  & $3.011\times10^{-2}/\left(124.40 - \sigma^{2}\right)$ & &  \\
    H$_2$O & $+ 7.46\times10^{-3}\cdot\left( 0.203 - \sigma \right)\cdot$ & {\citet{hill&lawrence1986}} & 0.36~$\mu$m < $\lambda$ < 19~$\mu$m \\
                  & $\left(1.03-1.98\times10^3\sigma^2+8.1\times10^4\sigma^4-1.7\times10^8\sigma^8\right)^{-1}$ & & \\
    \hline
  \end{tabular}
$^{1}$See also \citet{mikhail_refract}.
\end{table}
\clearpage
\begin{table}[ht]
  \scriptsize
  { Table 2. Refractive Boundary Pressures for Isothermal Jupiters} \\
  \vspace{2mm}
  \begin{tabular}{c c c c c c c c c c c c}
    \hline
    \hline
                           &                     &                   &     300 K      &             &           &   600 K       &            &            &  1200 K      &      \\
    Host Star        &  $T_{\rm{s}}$& B16$^{1}$ &   Eq. 14       &  frac.   &   B16 &   Eq. 14      &  frac.   &   B16   &   Eq. 14      &  frac.    \\
    Spectral Type &           (K)      & (bar)          &   (bar)        &    error   &   (bar) &   (bar)       &  error       &   (bar) &  (bar)       &  error    \\
    \hline
    F0                  &  7300             &   0.37       &   0.39       &  0.05        &   3.97   &   4.39     &   0.11     &    38.5     &    49.7     &  0.29  \\
    G2                 &  5778             &    0.60       &   0.64       &  0.07        &   6.20   &  7.23     &   0.17     &   54.3      &    81.8     &  0.51  \\
    K5                  & 4410             &    1.02       &   1.14       &  0.11        &   9.90   &   12.9    &    0.30     &    72.3    &     145      &  1.01  \\
    M2                  & 3400            &    1.73       &   2.06       &   0.19        &   14.7   &  23.3      &  0.58      &    83.5     &    264     &  2.16  \\
    \hline
  \end{tabular}
  
$^{1}$From \citet{betremieux2016}, their Table~3.
\end{table}
\clearpage
\clearpage
\clearpage
\clearpage
\clearpage
\clearpage
%
\end{document}